\documentclass[aps,prc,twocolumn,groupedaddress,showpacs]{revtex4}
\usepackage{graphicx}
\usepackage{amsmath}

\usepackage[colorlinks,citecolor=blue,urlcolor=blue,linkcolor=blue]{hyperref}

\def\eqref#1{Eq.~(\ref{eq:#1})}
\def\eqlab#1{\label{eq:#1}}
\def\figref#1{Fig.~(\ref{fig:#1})}
\def\figlab#1{\label{fig:#1}}
\def\tabref#1{Tab.~(\ref{tab:#1})}
\def\tablab#1{\label{tab:#1}}
\def\seclab#1{\label{sect:#1}}
\def\secref#1{Section~\ref{sect:#1}}

\begin{document}


\title{Equation of state of nuclear matter from collective flows and stopping in intermediate energy heavy-ion collisions}


\author{M.D. Cozma}
\email{dan.cozma@theory.nipne.ro}
\affiliation{Department of Theoretical Physics, IFIN-HH,\\ Reactorului 30, 077125 M\v{a}gurele/Bucharest, Romania}


\date{\today}

\begin{abstract}
The equation of state of nuclear matter, momentum dependence of the effective interaction and in-medium modification of elastic nucleon-nucleon cross-sections are studied by comparing theoretical predictions for stopping, directed and elliptic flows of protons and light clusters in intermediate energy heavy-ion collisions of beam energy between 150 and 800 MeV/nucleon to experimental data gathered by the FOPI Collaboration. To that end, the dcQMD transport model was upgraded by implementing medium modifications of differential elastic cross-sections guided by microscopical model calculations, a medium modification factor of elastic transition amplitudes that depends on the local density, isospin asymmetry and isospin projection of the scattering particles, a MST coalescence algorithm applied at the local freeze-out rather than at the end of simulation and threshold effects for elastic scattering. A multivariate analysis that takes into account systematic uncertainties induced on model predictions by the coalescence afterburner leads to the following constraint for the equation of state at 68 percent confidence level: compressibility modulus of isospin symmetric matter $K_0=230^{+9}_{-11}$ MeV and slope of the symmetry energy $L=63^{+10}_{-13}$ MeV. The momentum dependence of the isoscalar potential is found to be similar to that of the empirical optical potential, with an effective isoscalar mass $m^*=0.695^{+0.014}_{-0.018}$. The isovector potential displays a momentum dependence corresponding to a positive neutron-proton effective mass difference $\Delta m^*_{np}=(0.17^{+0.10}_{-0.09})\delta$, close to the world average for this quantity. A suppression of elastic nucleon-nucleon cross-sections in symmetric nuclear matter at saturation by about 60$\%$ relative to vacuum values is deduced, in qualitative agreement with microscopical results. A strong dependence of the suppression factor on isospin asymmetry is evidenced, experimental data for isospin symmetric systems proving crucial for this last conclusion.

\end{abstract}


\maketitle

\section{Introduction}

Heavy-ion collisions (HIC) at intermediate energy provide the opportunity to study different aspects of the in-medium effective nucleon-nucleon interaction: in-medium elastic nucleon-nucleon cross-sections~\cite{Gaitanos:2004ic,Prassa:2005ag,Zhang:2007gd,Basrak:2016cbo,Barker:2016hqv,Wang:2020xgk}, momentum dependence of the optical potential~\cite{Aichelin:1987ti,Pan:1992ef,Danielewicz:1999zn,Larionov:2000cu,Coupland:2014gya,Malik:2018juj,Morfouace:2019jky,Nara:2020ztb,SpRIT:2023als} and equation of state (EoS) of both symmetric (SNM)~\cite{Danielewicz:2002pu,Ivanov:2014ioa,Pratt:2015zsa,LeFevre:2015paj,Nara:2018ijw,Oliinychenko:2022uvy,Steinheimer:2022gqb} and asymmetric nuclear matter (ANM)~\cite{Chen:2004si,Tsang:2004zz,Tsang:2008ua,Russotto:2011hq,Wang:2014rva,Russotto:2016ucm,Wang:2020dru,Cozma:2017bre,SpiRIT:2021gtq}. The last topic, in particular the isovector component of the EoS, commonly known as the symmetry energy (SE), has attracted much attention during the last decade especially since the advent of multi-messenger astronomy~\cite{LIGOScientific:2017ync,LIGOScientific:2017vwq} which has offered the opportunity of measuring properties of neutron stars with greater accuracy~\cite{Raaijmakers:2021uju,Miller:2021qha} and opened the possibility of answering questions related to the nature of matter at the core of such compact objects~\cite{Bombaci:2004mt,Drago:2013fsa,Bombaci:2020vgw,Tang:2021snt}, in particular finding robust solutions to the long-standing hyperon puzzle~\cite{Djapo:2008au,Bombaci:2016xzl,Miyatsu:2013yta,Chorozidou:2024gyy}.

The interpretation of HIC observables measured in the laboratory relies most often on transport models~\cite{Feldmeier:1989st,Danielewicz:1991dh,Ono:1998yd,Hartnack:1997ez,Buss:2011mx,TMEP:2022xjg} which contain the three above mentioned quantities as ingredients. These represent three different facets of the same underlying quantity, namely the in-medium nucleon-nucleon interaction, and can in principle be determined using microscopic ab-initio models starting from the relatively well known two-nucleon and three-nucleon interactions in vacuum~\cite{Wellenhofer:2014hya,Holt:2015dfa,Keller:2022crb} and used as input in transport models. This is a daunting task, since HIC probe a wide range of densities, isospin asymmetries and temperature or more generally non-equilibrium states~\cite{Gaitanos:2004va,Xu:2019ouo}. While such approaches are desirable and promising attempts in this direction have been made~\cite{Gaitanos:2001hv,Gaitanos:2004ic}, the most often employed technique has been of the divide-and-conquer type~\cite{Danielewicz:1999zn,Danielewicz:2002pu,Russotto:2011hq}. Identifying observables that are sensitive to a particular model ingredient while being relatively insensitive to others has made possible conclusions most often at qualitative level~\cite{FOPI:2004bfz,Fuchs:2000kp,Hartnack:2005tr} and only rarely at a quantitative one~\cite{Danielewicz:2002pu,LeFevre:2015paj,Russotto:2011hq,Russotto:2016ucm,Cozma:2017bre,SpiRIT:2021gtq}.

A divide-and-conquer approach is applicable if no correlations among transport model ingredients exist. This is however valid only at most as a first approximation. Correlations between in-medium cross-sections and the optical potential emerge at fundamental level as aspects of the same in-medium corrections of the $NN$ interaction encoded in nucleon self-energies. At higher level, effective masses, which characterize the strength of the momentum dependence of the optical potential, enter in the expression of in-medium cross-sections~\cite{Li:2005jy,Larionov:2003av,Larionov:2007hy}. Empirically, correlations between the momentum dependence of the optical potential and compressibility modulus of SNM have been observed once theoretical predictions have been compared to HIC experimental data~\cite{Tarasovicova:2024isp}. Generally, a hierarchy of in-medium effects on HIC observables at intermediate energies can be evidenced: in-medium corrections of cross-sections have the strongest impact, followed by sensitivity to momentum dependence of the optical potential, with EoS having the smallest effect. Residual dependence on the first two ingredients can still have a noticeable impact even for the most carefully selected observables for EoS related studies. This issue has been most often circumvented by using information from sources other than HIC (e.g. isoscalar component of the optical potential~\cite{Aichelin:1991xy,Hartnack:1994zz}). Nevertheless, a possible dependence of in-medium cross-sections on isospin asymmetry has never been considered in HIC studies. 

Two main categories of observables can be accessed experimentally in intermediate energy HIC: nucleonic and particle production. Nucleonic rapidity distributions, in particular stopping observables such as $varxz$ (or $vartl$ in earlier studies)~\cite{FOPI:2010xrt}, are known to be mainly sensitive to in-medium cross-sections, though a more careful analysis reveals also a rather loose correlation between the suppression factor of cross-sections and nucleon effective masses and some residual dependence on the EoS. Collective flows, in particular the slope of directed flow at mid-rapidity and rapidity or transverse momentum dependent elliptic flow, have been traditionally used to study EoS of both SNM~\cite{Danielewicz:2002pu,LeFevre:2015paj} and ANM~\cite{Russotto:2011hq,Russotto:2016ucm,Cozma:2017bre}, but also effective masses~\cite{Danielewicz:1999zn} and indirectly in-medium cross-sections due to correlations between directed flow and stopping~\cite{FOPI:2004orn,Zhang:2007gd}. 

Particle production studies were pursued as means to access more exclusive information about the fireball formed during HIC in relation to EoS studies above saturation. In particular, kaon production appeared particularly promising due to the weak final-state interaction with the dense nuclear medium and the magnifying glass effect of emission below the vacuum threshold~\cite{Aichelin:1985rbt}. Measurement of $K^+$ production in C+C and Au+Au at 800 MeV/nucleon impact energy~\cite{KAOS:2000ekm} has allowed the extraction of a soft constraint for EoS of SNM~\cite{Fuchs:2000kp,Hartnack:2005tr}. Attempts to study the symmetry energy with kaon production has revealed a lack of sensitivity in HIC relative to infinite nuclear matter~\cite{Ferini:2006je}, yet to be understood in detail, and insufficient experimental accuracy~\cite{FOPI:2007gvb}. Pions are more copiously produced at intermediate energies and both charged partners of the isovector triplet are easily accessible experimentally. Consequently, the charged pion multiplicity ratio has been proposed for study of the symmetry energy~\cite{Li:2004cq} but has initially lead to significant confusion and contradictory results. It has been later realized that inclusion of threshold effects~\cite{Ferini:2006je,Song:2015hua,Cozma:2014yna} in the collision term are required for a thermodynamically consistent description~\cite{Zhang:2017nck}. A successful description of S$\pi$RIT pion production data below threshold (Sn+Sn at 270 MeV/nucleon) was possible~\cite{SpiRIT:2021gtq} once additional difficulties related to its sensitivity to baryonic resonance potentials and the momentum dependence of the nucleon Lane potential~\cite{Cozma:2021tfu,Ikeno:2023cyc} were understood and surmounted.


A large database of experimental data for HIC observables at intermediate energies for systems with different isospin content over a wide range of impact energies has been accumulated, most notably by the FOPI~\cite{FOPI:2003fyz,FOPI:2006ifg,FOPI:2010xrt,FOPI:2011aa}, HADES~\cite{HADES:2020lob,HADES:2020ver}, S$\pi$RIT~\cite{SpiRIT:2021gtq,SpRIT:2023als}, KaOS~\cite{KAOS:2000ekm,Forster:2007qk} and a few other collaborations~\cite{FOPI:1993wdf,Lukasik:2004df,Coupland:2014gya,Russotto:2016ucm}. Additional systematic measurements, in particular scans in impact energy, will be carried out in the near future by ASYEOS2~\cite{Russotto:2021mpu} (AuAu at 250,400 and 800 MeV/nucleon) and HADES~\cite{Harabasz:2023rwc} (AuAu at 400, 600 and 800 MeV/nucleon) Collaborations at GSI, but also S$\pi$RIT Collaboration at RIKEN. On a longer timescale, measurements performed at FRIB and its proposed energy upgrade FRIB400~\cite{FRIB400:2019aaa} have the potential to provide additional valuable information.

Time is ripe to make use of this wealth of HIC experimental data to their full potential and extract quantitative information on all three aspects of the in-medium $NN$ interaction. To that end, a transport model that can reliably describe both nucleonic and particle production observables over a wide range of beam energies is required. Ultimately, an estimate of the residual model dependence, i.e. variation of results with the transport model used, should also be determined. This last goal necessitates a significant community effort and is actively pursued by several groups~\cite{Kolomeitsev:2004np,Reichert:2021ljd,Zhang:2017esm,Ono:2019ndq,Colonna:2021xuh,TMEP:2023ifw}. 

The present study aims at representing a first step towards determining accurate information on the EoS of nuclear matter above saturation density and other aspects of the effective in-medium nucleon-nucleon interaction from HIC. The complexity of the analysis is kept close to the lowest possible by restricting description to only nucleonic observables over a narrower range of beam energies than currently available. The main motivation for such a limitation is to avoid systems for which baryonic resonance degrees a freedom have an impact on time evolution of the reaction. Using the transport model of choice it was found that at impact energies up to 800 MeV/nucleon the fraction of nucleons excited into baryonic resonances, mainly $\Delta$(1232), never exceeds 5$\%$ in mid-central HIC, in agreement with similar estimates~\cite{Ehehalt:1993cx}. Varying the $\Delta(1232)$ potential in nuclear matter can have an impact on nucleonic flows in Au+Au collisions at impact energies of 1.0 GeV/nucleon comparable or even larger than experimental accuracy. Fixing these quantities is most reliably achieved by describing pion production~\cite{Cozma:2021tfu,Ikeno:2023cyc}. An extension of the current study in this direction is postponed for a later date in view of an existing tension between FOPI and HADES experimental data for these observables~\cite{HADES:2020ver} and needed improvements of the transport model to robustly describe reactions above 1.0 GeV/nucleon beam energy, see \secref{eos}.

The FOPI Collaboration has performed systematical measurements of HIC at beam energies from 0.09 to 1.5 GeV/nucleon. A large fraction of the available data set consists of nucleonic observables such a stopping~\cite{FOPI:2004orn,FOPI:2010xrt}, directed ($v_1$) and elliptic ($v_2$) flows~\cite{FOPI:2003fyz,FOPI:2004bfz,FOPI:2011aa}. Results for particle production, in particular pions~\cite{FOPI:2006ifg}, but also kaons, $\phi$ meson and other strangeness carrying hadrons are also available. The FOPI database provides the most extensive collection of experimental measurements of nucleonic observables at intermediate impact energy and therefore constitutes a natural first choice for EoS related investigations. The only quantitative analysis to date employing it~\cite{LeFevre:2015paj} has made use only of rapidity dependent elliptic flow data between 0.4 and 1.5 GeV/nucleon concluding that $K_0$=190$\pm$30 MeV at 68$\%$ CL. Directed flow is an equally viable observable for investigations of the EoS of SNM, however an outstanding tension between constraints using directed and elliptic flow exists~\cite{Danielewicz:2002pu,Oliinychenko:2022uvy}. In the seminal work of Danielewicz $et\,al.$~\cite{Danielewicz:2002pu} the slope of directed flow at mid-rapidity favors a soft EoS with $K_0$ in the range 167-210 MeV, while experimental data for $v_2$, overwhelmingly for reactions above 1.0 GeV/nucleon beam energy, lend support to a stiffer EoS, with $K_0\approx$ 300 MeV. 

Experimental data for both $v_1$ and $v_2$ at reduced mid-rapidity, $|y/y_P|\leq 0.5$, are considered in the present work. The restriction on rapidity was imposed since theoretical transverse momentum dependent  proton direct flow spectra away from mid-rapidity were found to be contaminated by misidentified free protons, see Appendix A. Results for the stopping observable $varxz$ are also employed, as they provide strong constraints for the in-medium modification factor of cross-sections~\cite{Danielewicz:1999zn}. Given the availability of data for both isospin symmetric and neutron rich systems, $varxz$ facilitates breaking the degeneracy between isospin asymmetry independent and dependent in-medium modification factors of elastic cross-sections that would otherwise occur if only data for isospin asymmetric (mostly $^{197}$Au+$^{197}$Au) were included. Directed flow data for $^{58}$Ni+$^{58}$Ni play a similar role.

The present study is a loose extension of Ref.~\cite{Cozma:2017bre}, the latest version of the dcQMD transport model~\cite{Cozma:2021tfu} is however used and upgraded. General features of the model as well as details related to determination of in-medium elastic $NN$ cross-sections, modified Pauli blocking and MST coalescence algorithms are presented in~\secref{themodel}. The main results of the paper are discussed in \secref{results}: the experimental data set and computation algorithm are detailed, followed by a presentation of constraints on model parameters extracted from a fit to experimental data for $v_1$, $v_2$ and $varxz$. Special emphasis is put on highlighting the importance of threshold effects, applying coalescence at the local freeze-out time and isospin asymmetry dependence of in-medium cross-sections.~\secref{discussion} is devoted to a discussion of the implications of describing the chosen experimental data set. It starts with a quantitative estimation of the probed density range and associated sensitivity, in~\secref{probeddens}. The latter part of \secref{discussion} consists of a presentation of the extracted constraints for the EoS of both SNM and ANM in~\secref{eos}, in-medium elastic $NN$ cross-sections in~\secref{inmedcs} and momentum dependence of the optical potential in~\secref{mdi}. Comparisons to microscopical model results and other empirical studies are included. A section devoted to summary and conclusions follows,~\secref{sumcon}. Model predictions are compared to the fitted experimental data set and several omitted observables in Appendix A. 

\section{The model}
\seclab{themodel}
\subsection{The Transport Model}
\seclab{transpmodel}
Heavy-ion collision dynamics is simulated using an upgraded version of the dcQMD quantum molecular dynamics (QMD) model~\cite{Cozma:2021tfu}. The general framework of semi-classical description of heavy-ion collisions is briefly described in the following, while most of the section is devoted to a full account of upgrades.

QMD transport models provide an approximate solution to the time-dependent Schr\"odinger equation describing a system of N particles. By assuming the total wave-function of the system to be the product of single-particle wave-functions (Hartree approximation), it can be shown that the time dependence of expectation values of single particle operators is described by the classical Hamilton equations of motion~\cite{deGroot:1972aa,Hartnack:1997ez}. For the position and momentum operators they read
\begin{eqnarray}
\eqlab{eomqmd}
\frac{d\vec{r}_i}{dt}=\frac{\partial \langle U_i \rangle}{\partial \vec{p}_i}+\frac{\vec{p}_i}{m},\qquad
\frac{d\vec{p}_i}{dt}=-\frac{\partial \langle U_i \rangle}{\partial \vec{r}_i}\,.
\end{eqnarray}
The average of the potential operator of particle $i$ is determined by averaging its Weyl transform, weighted by the Wigner function, over the entire phase-space~\cite{deGroot:1972aa}. In this work, the potential $U_i$ is the sum of the Coulomb and strong potentials. The latter carries explicit dependence on density $\rho$, isospin asymmetry $\delta$ and momentum. The full expression of the classical Hamiltonian to be used in the above equations of motion reads
\begin{eqnarray}
\eqlab{hamiltonian}
 \langle H \rangle&=&\sum_i \sqrt{p_i^2+m_i^2}+\sum_{i,j,j>i}\,\bigg[\frac{A_+}{2}+\tilde\tau_i\,\tilde\tau_j\,\frac{A_-}{2}\bigg]\,u_{ij} \nonumber\\
 &+&\sum_{i,j,j>i}\bigg[C_+ +\tilde\tau_i\,\tilde\tau_j\,C_-\bigg]\frac{u_{ij}}{1+(\vec{p_i}-\vec{p_j})^2/\Lambda^2}\nonumber\\
 &+&\sum_i\,\frac{B}{\sigma +1}\,[1-x\tilde\tau_i\,\delta_i\,]\,u_i^\sigma+\frac{D}{\eta+1}\,[1-y\tilde\tau_i\,\delta_i]\,u_i^\eta \nonumber\\
 &+&\sum_{i,j,j>i} U_{ij}^{Coul}
\end{eqnarray}
where $\tilde\tau_i$=-$\tau_i/T_i$, $u_{ij}=\rho_{ij}/\rho_0$ is the partial relative interaction density \cite{Hartnack:1997ez,Aichelin:1988me} of particles $i$ and $j$ with $u_i=\sum_{j\neq i} u_{ij}$. Here $T_i$ and $\tau_i$ denote the isospin and isospin projection of particle $i$ respectively. In this study we require that the isoscalar and isovector resonance potentials (only $\Delta(1232)$ for the considered range of impact energy) are equal to those of the nucleon, a simplification in comparison to Ref.~\cite{Cozma:2021tfu}. Consequently, the 11 free parameters in the expression above do not carry an index denoting particle specie. They and an additional one (density $\tilde{\rho}$ at which the magnitude of symmetry energy is fixed) can be thus determined by requiring that certain properties of EoS of cold nuclear matter and momentum dependence of the interaction be described~\cite{Cozma:2017bre}. To make the comparison with Ref.~\cite{Cozma:2017bre} completely transparent, shorthand notations $A_{+/-}=A_l\pm A_u$ and $C_{+/-}=C_l\pm C_u$ were used in the expression above. Additionally, a new parameter $\eta$ has been introduced to enforce values larger than 1.0, whenever possible, for $\sigma$, which provides certain technical advantages in the numerical implementation of the model.

The scattering term includes elastic and inelastic two-baryon collisions ($N+N\rightarrow N+N$, $N+N\rightarrow N+R$, $N+R\rightarrow N+R'$, etc.), resonance decays into a pion-nucleon pair ($R\rightarrow N+\pi$ ) and single pion absorption reactions ($\pi+N\rightarrow R$). Imposing energy conservation at the level of this term induces threshold effects~\cite{Ferini:2006je,Song:2015hua,Cozma:2014yna,Zhang:2017mps,Ikeno:2023cyc}, which are required for thermodynamic consistency~\cite{Zhang:2017nck}. This implies that final state kinematics of a two-body collision, resonance decay or pion absorption event is determined by taking into account the in-medium potential energy of all particles in the system. Modification of in-medium cross-sections for inelastic channels is treated as described in Ref.~\cite{Cozma:2021tfu}. For elastic nucleon-nucleon scattering the approach in that reference is altered---, as described below, to resemble the one used for inelastic channels more closely.

\subsection {In-medium elastic nucleon-nucleon cross-sections}
\seclab{inemedelcs}

The definition for differential cross-section of two-particles scattering into a n-particles final state, in vacuum, reads~\cite{weinberg::1996aa}
\begin{eqnarray}
 d\sigma=(2\pi)^4 u_i^{-1} |M_{fi}|^2 \delta^4(\sum_i p_i-\sum_f p_f) \prod_f d^3 p_f
 \eqlab{csvacdef}
\end{eqnarray}
where $u_{i}$ represents the relative velocity of the two colliding particles, while $M_{fi}$ is the transition amplitude. For the case of scattering of spin 1/2 particles, a final state of multiplicity equal to 2 and using a spinor normalization that leads to an expression for these quantities with a leading term equal to one in the static limit~\cite{bjorkendrell::1964aa}, the above expression can be simplified to read
\begin{eqnarray}
 \frac{d\sigma}{d\Omega}=(2\pi)^4 \frac{m_1 \,m_2}{k_i \sqrt{s_i}}|M_{fi}|^2 \frac{k_f\,m_{1'}\,m_{2'}}{\sqrt{s_f}}.
 \eqlab{cs2bvac}
\end{eqnarray}
Here $m_{1,2}$ and $m_{1',2'}$ represent the rest masses of the particles in the initial and final states respectively. The labels $i$ and $f$ attached to the relative momentum $k$ and invariant mass $s$ have the same meaning. The expression was written in a form that anticipates the introduction of medium modifications, however in vacuum the equality $s_i$=$s_f$ does hold.

In-medium cross-sections can be defined in a similar manner, with the four-momentum conservation in~\eqref{csvacdef} replaced by in-medium kinetic momentum and in-medium single particle energy conservation. The model described by the effective Hamiltonian ~\eqref{hamiltonian} contains only scalar interactions which implies that kinetic and canonic momenta are the same. In addition, conservation of the total energy of the system is imposed, which in general is not equivalent to conservation of single particle energies of the scattering entities. Consequently, the following expression for the in-medium differential cross-section can be derived 
\begin{eqnarray}
 \frac{d\sigma}{d\Omega}&=&(2\pi)^4 \frac{m_1^* \,m_2^*}{k_i^* \sqrt{s_i^*}} |M_{fi}^{(med)}(\rho,\delta,\{\tau\})|^2\frac{k_f^*\,m_{1'}^*\,m_{2'}^*}{\sqrt{s_f^*}}. 
 \eqlab{cs2bmed}
 \end{eqnarray}
 It is identical in form with ~\eqref{cs2bvac}, the only differences being that vacuum masses are replaced by in-medium ones everywhere in the kinematic factor (also in the expression of the relative momentum $k$ and invariant mass $s$) and the transition amplitude becomes a quantity that can in principle depend on additional variables compared to vacuum. The former are determined using the equations of motion \eqref{eomqmd} and thus depend on local density, isospin asymmetry, isospin projection of the particle and local momentum distribution (or temperature in case of local equilibrium). For the latter we only keep and explicit dependence on density, isospin asymmetry and projection of the total isospin. A dependence on the modulus of each particle's momentum in the local rest-frame of nuclear matter is in principle also possible (for the case of an infinite isotropic system, for the case of heavy-ion collisions the situation is more involved). We nevertheless make the simplifying assumption that the momentum dependence of medium-modifications of cross-sections is fully captured through the appearance of effective masses in ~\eqref{cs2bmed}. For the medium modified scattering amplitude the following Ansatz is used
\begin{eqnarray}   
 |M_{fi}^{(med)}(\rho,\delta,\{\tau\})|^2 = \frac{1}{2} ( |M_{fi}^{(vac)}(\tilde{s}_i)|^2+|M_{fi}^{(vac)}(\tilde{s}_f)|^2)\nonumber\\
 \times\exp{[(\alpha+\beta_1\,\delta+\beta_2\,(\tau_1+\tau_2)\delta)\frac{\rho}{\rho_0}]}. 
 \eqlab{sameddef}
\end{eqnarray}
At low densities this induces a medium modification factor of cross-sections that depends linearly on density, in agreement
with the low density theorem~\cite{ericsonweise::1988aa} and several empirical determinations of in-medium elastic nucleon-nucleon cross-sections~\cite{Zhang:2007gd,Basrak:2016cbo,Wang:2020xgk}. $M_{fi}^{(med)}$ in the expression above is written in terms of the vacuum scattering amplitude $M_{fi}^{(vac)}$ evaluated at two different effective invariant masses $\sqrt{\tilde{s}_i}$ and $\sqrt{\tilde{s}_f}$, that are defined such that they are above the vacuum threshold by the same magnitude as $\sqrt{s_i^*}$ and $\sqrt{s_f^*}$ are above the in-medium threshold,
\begin{eqnarray}
 \sqrt{\tilde{s}_{i,f}}-2\,m_N= \sqrt{s_{i,f}^*}-\sqrt{s_{th}^*}+U_{i,f}-U_{th}.
\end{eqnarray}
The value of the threshold invariant mass $s_{th}^*$ is determined by the requirement that the relative momentum of the scattered particles is zero at constant total momentum. The last two terms in the expression above account for an effect due the momentum dependence of the potential energy. The nucleon mass in vacuum is denoted by $m_N$.

At finite density the vacuum transition amplitude is dressed by medium corrections. To compute these, a microscopical model is in principle needed. In a Feynman diagramatic picture this amounts to adding medium corrections to the initial, intermediate and final states of the vacuum transition amplitude. The total transition amplitude can then be computed as the coherent sum of these three contributions. If the intermediate state contribution and the interference terms are neglected the final result for $|M_{fi}^{(med)}|^2$ is the incoherent sum of the squares of medium corrected vacuum transition amplitude in the initial and final states, with the medium correction only modifying the value of the invariant mass. This approach only takes into account effects induced by energy and momentum conservation in the medium. Additional medium effects are in principle possible and are simulated by allowing an explicit dependence of the transition amplitude on density, isospin asymmetry and total isospin projection, as introduced by the multiplicative factor in~\eqref{sameddef}. The a priori unknown parameters $\alpha, \beta_1, \beta_2$ will be determined from a comparison to experimental data. They induce a density, isospin asymmetry and isospin splitting of elastic nucleon-nucleon cross-sections on top of similar effects induced by in-medium conservation of energy and momentum and by effective masses.

\begin{figure}
 \includegraphics[width=0.485\textwidth]{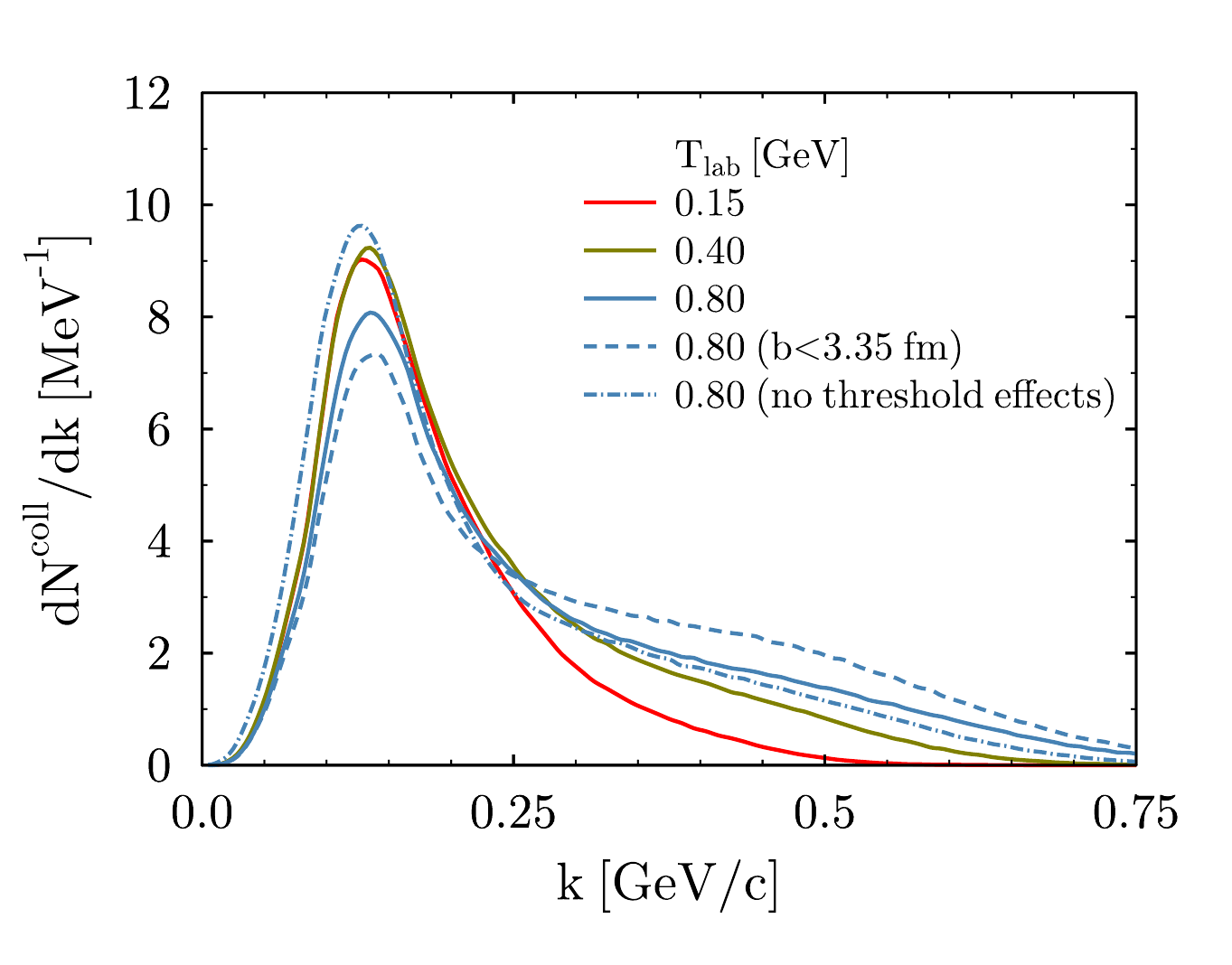}
 \caption{\figlab{collstat} (Color online) Distribution of successful two-body $\it{elastic}$ collisions in mid-central, unless otherwise specified, AuAu collisions as a function of the relative initial state momentum $k$. The importance of threshold effects and change of the centrality range is shown for $T_{lab}$=0.80 GeV/nucleon.}
\end{figure}

Total cross-sections can be obtained by integrating ~\eqref{cs2bmed} over the solid angle. For the type of interaction chosen in~\eqref{hamiltonian} and ranges in parameter set used in this study, the angular dependence induced by in-medium correction in the flux and phase-space factor are moderate. Consequently, the same expression can be written for integrated cross-sections, with the solid angle dependent transition amplitude replaced by an integrated one that can be expressed in term of the corresponding vacuum quantities using an identical expression as in~\eqref{sameddef} and the flux and phase-space factor evaluated for an arbitrary orientation of the final state relative momentum. In practice, the orientation provided by in-medium differential cross-section distributions is used. The needed vacuum integrated transition amplitudes are determined using the expression for total vacuum cross-sections and the parametrizations below for vacuum elastic $np$ and $nn/pp$ cross-sections
\begin{widetext}
\begin{displaymath}
 \sigma_{np}^{(vac)} = \left\{ \begin{array}{ll}
790,&\quad\textrm{$p\leq0.15$}\\
91.00/(p-0.0524)-82.6-371.05\,p &\quad\textrm{$0.15 < p \leq0.25$}\\
48.838/(p-0.1278)-173.46+274.94\,p-156.08\textrm{\,$p^{2}$}+31.18\textrm{\,$p^{3}$}, &\quad\textrm{$0.25 < p \leq2.0$}\\
77.0/(p + 1.5), &\quad\textrm{$p > 2.0$}
\end{array} \right.
\end{displaymath}
\begin{equation*}
 \sigma_{nn,pp}^{(vac)}= \left\{ 
 \begin{array}{ll}
260.0,& \quad\textrm{$p\leq0.15$}\\
38.81/(p-0.0520)-181.56+327.22\,p - 169.86\textrm{\,$p^2$},&\quad \textrm{$0.14 < p \leq0.8$}\\ 
174.13+559.68\,p-735.73\textrm{\,$p^2$}-401.02\textrm{\,$p^3$}+77.23\textrm{\,$p^4$},&\quad\textrm{$0.8 < p \leq2.0$}\\
77.0/(p + 1.5), &\quad\textrm{$p > 2.0$}
\end{array} \right.
\end{equation*}
\end{widetext}

\begin{figure*}[htb]
\includegraphics[width=0.495\textwidth]{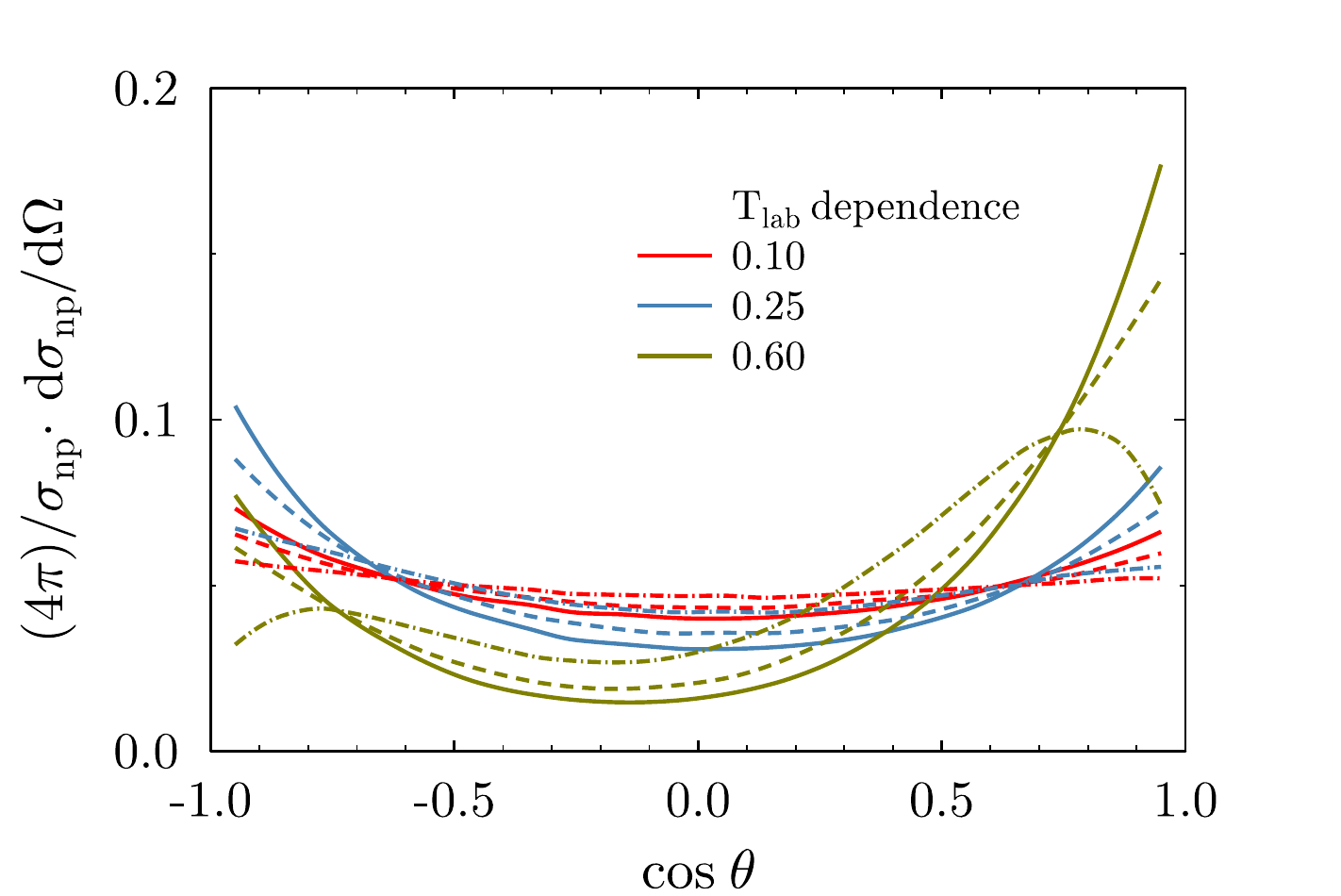}
\includegraphics[width=0.495\textwidth]{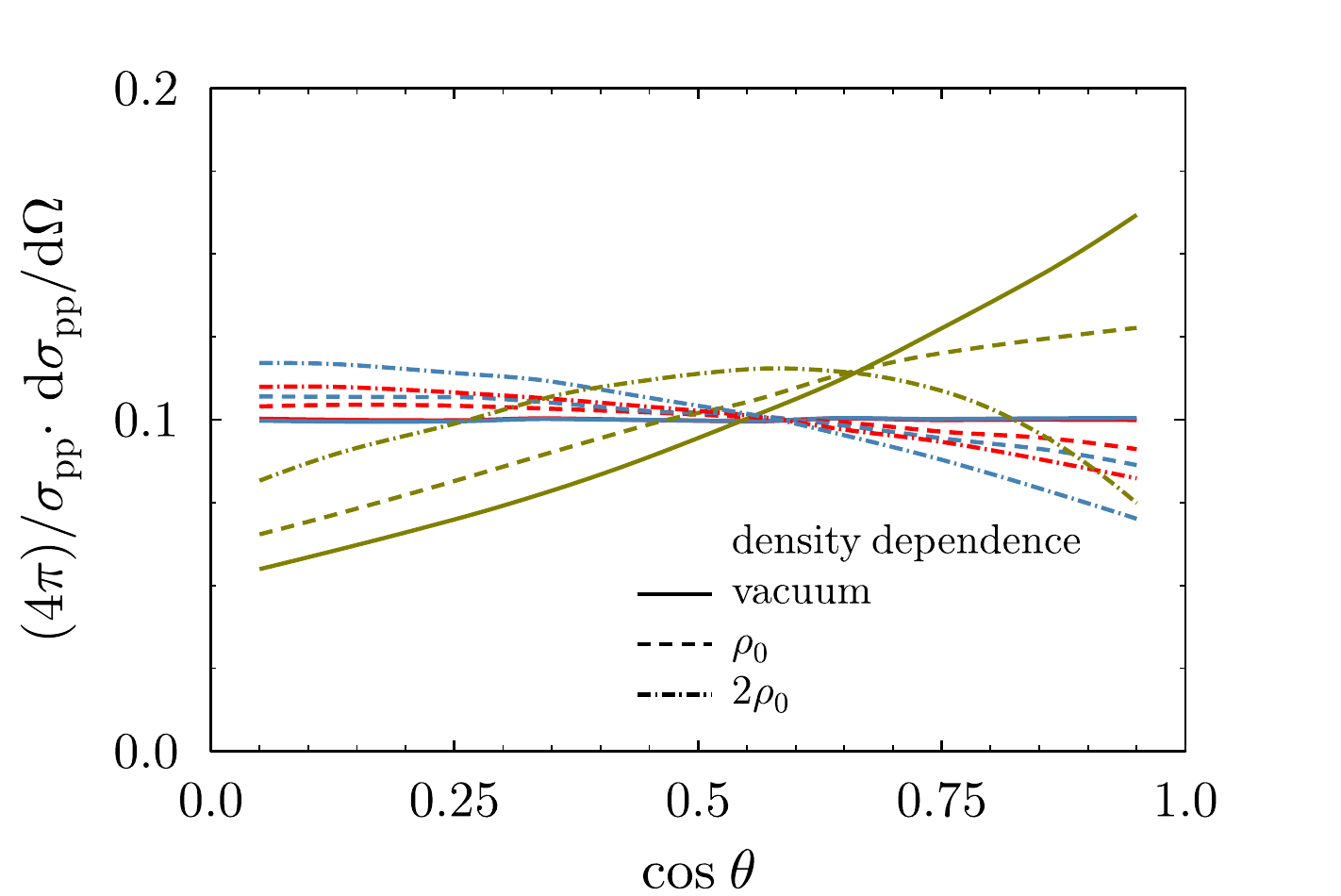}
\caption{\figlab{meddifcs} (Color online) Density dependence of in-medium $np$ (left panel) and $pp$ (right panel) differential elastic cross-sections. Numerical results of used parametrization are shown for combinations of three impact kinetic energies, $T_{lab}$=0.10, 0.25 and 0.60 GeV/nucleon (red, blue and green curves respectively), and three values for density, $\rho=0, \rho_0, 2\rho_0$ (full, dashed and dash-dotted curves respectively).}
\end{figure*}

In the above expressions $p$ represents the projectile momentum in the laboratory frame, expressed in units of [GeV/c], while cross-section values are determined in units of mb. The above parametrizations are identical to those introduced by Cugnon et al.~\cite{Cugnon:1996kh} for $p \geq$ 2.0 GeV/c. They have been adjusted below this threshold to reproduce the latest experimental data set \cite{ParticleDataGroup:2018ovx}, in particular below 200 MeV. In this region, the original Cugnon parametrization represents effective in-medium cross-sections, without an explicit density or isospin asymmetry dependence but displaying an isospin dependence (as a result of the isospin dependence of vacuum values). This approach has the potential of introducing an unrealistic bias. Due to the rapid increase of vacuum elastic cross-sections towards lower energies, in particular $\sigma_{np}^{(vac)}$, they are kept constant below a certain momentum $\tilde{p}$ which was chosen to be 0.15 GeV/c for both $np$ and $pp$ channels. Lowering $\tilde{p}$ below this value has only a small impact on the collision rate and its distribution as a function of the relative momentum $k$ even for HIC of the lowest impact energy used in this study, $T_{lab}$=150 MeV/nucleon. A higher value has non-negligible impact on the extracted medium modification of elastic cross-sections, but the impact on other model parameters is within the 1-$\sigma$ range. Large values of cross-sections, as induced by the above parametrizations, lead to an increase of non-localities. However, for the magnitude of in-medium cross-sections deduced from fitting experimental data, see \secref{fitv1v2}, the increase is moderate, relative to that of Cugnon parametrization of cross-sections.

The distribution of two-body elastic collisions as function of the relative momentum $k$ in mid-central AuAu HIC for several impact energies (0.15,0.4,0.8 GeV/nucleon) is plotted in~\figref{collstat}. Besides a low energy peak common to all impact energies, a high energy tail, that becomes more prominent with increasing impact energy and extends beyond the $\Delta(1232)$ vacuum production threshold at $k=0.36$ GeV/c, is also observed. Switching off in-medium effects in the collision term leads to a visible increase of the collision rate at the low energy peak close to $k=0.15$ GeV/c which is the result of strong energy dependence of elastic $NN$ cross-sections in that region. About a quarter of collisions in the low energy peak take place above saturation density and about half of them below 0.5$\rho_0$, the latter accumulating at a uniform rate during the reaction. A similar analysis as performed in \secref{probeddens} reveals that flow observables are sensitive to the collision term evaluated at densities between 0.75 and 2.25 $\rho_0$. Consequently, collisions at very low density, of which an important fraction are spurious as result of Pauli blocking algorithm becoming less efficient with decreasing density, have no impact on the results of this study. The high energy tail displays a stronger beam energy dependence if inelastic collisions are added to the collision rate and a much stronger impact of threshold effects is observed at relative momenta close to the pion production threshold.

We close the discussion of in-medium elastic collisions by presenting the used angular distribution of final states. We use the Cugnon parametrization of differential cross-sections in vacuum. Their in-medium modification is taken to depend on density alone. The chosen parametrizations have been tuned to qualitatively agree with the microscopical calculations of Li and Machleidt~\cite{Li:1993ef,Li:1993rwa} for $pp$ and $np$ scattering up to the pion production threshold. With increasing density, differential cross-sections become more isotropic, a feature that is reached at densities above $\rho=2.0 \rho_0$. Extrapolations up to an impact energy of 800 MeV/nucleon attest a similar qualitative behavior. Since the algorithm of generating these distributions is not amenable to be cast as a simple parametrization, numerical results for several choices of density and impact energy are provided in \figref{meddifcs}. 

\begin{figure*}[htb]
\includegraphics[width=0.495\textwidth]{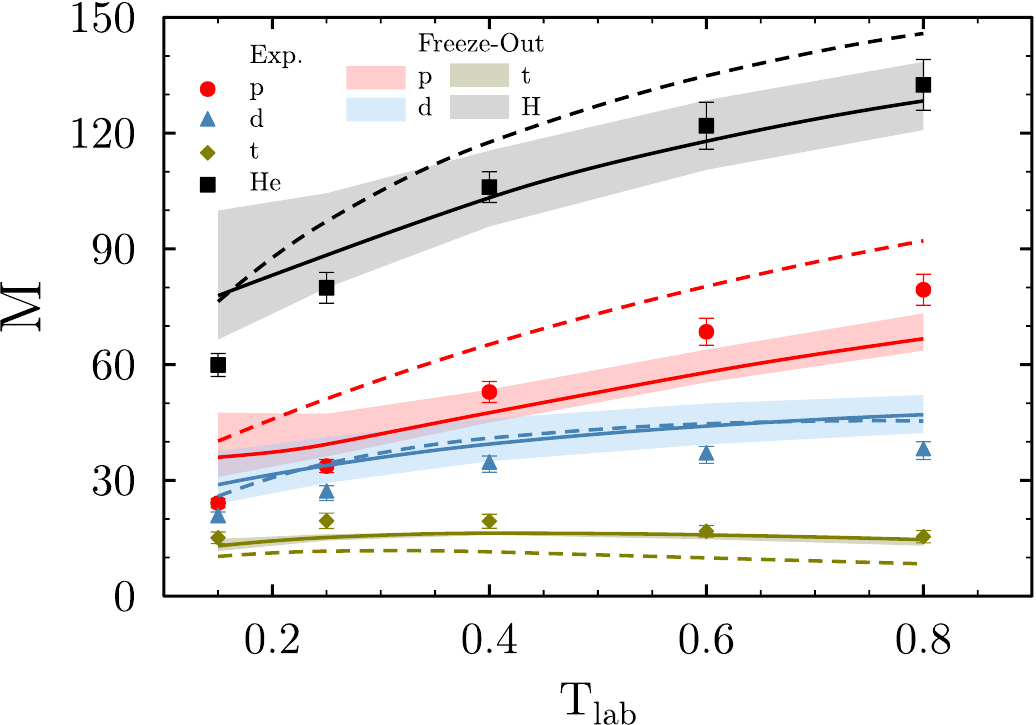}
\includegraphics[width=0.495\textwidth]{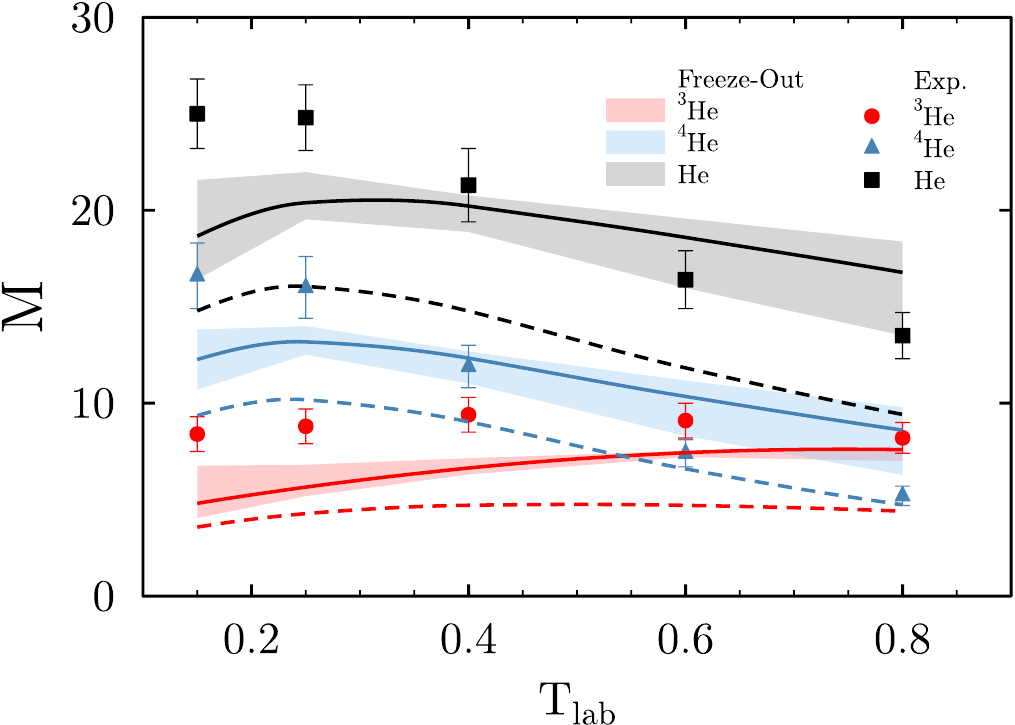}
\caption{\figlab{multcluster} (Color online) Comparison of theoretical values for cluster multiplicities for hydrogen (left panel) and helium (right panel) isotopes in central (b$<$2.0 fm) AuAu collisions with impact energy in the range 0.15-0.8 GeV/nucleon with experimental values (citation). The coalescence parameters were varied independently $\delta r$=3.0-4.0 fm and $\delta p$=0.2-0.3 GeV/c leading to a systematical uncertainty depicted as a band for each cluster specie. Experimental values~\cite{Reisdorf:2010aa} are depicted by symbols. Theoretical values for the choice $\delta r$=3.0 fm and $\delta p$=0.25 GeV/c are shown as full curves. Theoretical values determined by employing the coalescence afterburner at the end of the HIC simulation are shown as dashed curves for the choice $\delta r$=4.0 fm and $\delta p$=0.2 GeV/c.}
\end{figure*}

\subsection {Pauli blocking algorithm}
\seclab{pauliblocking}

Previous versions of the model~\cite{Cozma:2017bre,Cozma:2021tfu} made use of a Pauli blocking algorithm that included a surface correction term which increases the blocking probability of two-body collisions of nucleons that are close to the surface in either $r-$ or $p-$space~\cite{Aichelin:1991xy}. The correction was applied to nucleons with a single particle energy of less than 5 MeV, which induced a rather strong suppression of two-body collisions for densities below 3$\rho_0$. However, in heavy-ion collisions high density is correlated with temperature significantly above $T=0$ MeV and thus a strong surface correction term becomes unrealistic both in $r-$ and $p-$ space in the hot fireball. This strong suppression due to the surface term has also been evidenced recently in a comparison study of pion production in HIC~\cite{TMEP:2023ifw} as part of the Transport Model Evaluation Project (TMEP), where Pauli blocking of QMD models that use a surface correction term proved to be stronger than in BUU-type models. Consequently, in this study, the surface term is modified to only affect collisions in cold and moderately warm nuclear matter by only enforcing it for local densities below 1.25 $\rho_0$ and nucleon momentum lower than the local Fermi momentum. The latter requirement induces an isospin asymmetry dependence. Pion ratios remain almost unchanged, while total pion multiplicities increase by about 30$\%$ for SnSn collisions at 270 MeV/nucleon as compared to the values reported in ~\cite{TMEP:2023ifw}. The technical details of the (old) version of the surface correction term are presented in Ref.~\cite{Cozma:2017bre}.

\subsection{Coalescence afterburner}
\seclab{coalescence}

Final state spectra of heavy-ion collisions are determined by employing the minimum spanning-tree algorithm described in detail in Ref.~\cite{Cozma:2017bre}. In contrast to that study, the coalescence algorithm is applied at every time step during the reaction and identified clusters that do not undergo collisions with free nucleons or other clusters are considered to survive up to the end of the simulation. It should be noted that this approach merely represents a book-keeping algorithm that corrects for spurious nucleon evaporation that plagues QMD type models and has no impact on system dynamics. Consequently, formed clusters are identified at the local freeze-out time rather than at the end of the simulation (at t=150 fm/c in this study) which allows for smaller values of the $r$-space coalescence parameter $\delta r$=3.0-3.5 fm, as compared to the study in Ref.~\cite{Cozma:2017bre}. It has been previously noted that a very good description of experimental light cluster multiplicities can be achieved if unrealistically large values for $\delta r$ are adopted  ($\approx$10 fm) coupled with small values for $p$-space cut-off $\delta p$. An increase of the distance between two nucleons by 5 fm over a time span of 100 fm/c, that represents the duration between local freeze-out and end of simulation, can be achieved if the relative speed between nucleons is small relative to typical nucleon speeds ($\Delta p \approx$ 0.05 GeV/c). Consequently, a choice between values of $\delta r$ that differ by a few fm ($\leq$5 fm) induces a fine-tuning of coalescence parameters that cannot be easily justified. This offers support for an early identification of clusters at the local freeze-out time.

\begin{table*}[htb]
 \centering
\begin{tabular}{|c|c|l|}
\hline\hline
Parameter & Range & Explanation\\
\hline
$m^{*}$ & [0.6,0.9] & isoscalar effective mass \\
$U_\infty$ & [25.0,125.0] & strength of the isoscalar potential at $p\rightarrow\infty$\\
$K_0$ & [165.0,355.0] & compressibility modulus \\
$\alpha$ & [-0.4,0.8]  & density dependence of elastic scattering transition amplitude \\
& &in symmetric nuclear matter\\
\hline
$\Delta m^*_{np}/\delta$ & [-0.5,0.5] & neutron-proton effective mass splitting \\
$L$ & [15,145] & slope of density dependence of symmetry energy at saturation \\
$\beta_1$ & [-0.5,3.5] & isospin asymmetry dependence of transition amplitude \\
$\beta_2$ & [-1.5,2.5] & isospin splitting of $nn$ and $pp$ transition amplitudes \\
\hline\hline
\end{tabular}
\caption{List of tunable quantities and range of their variation during the fitting procedure. Parameters $\alpha$, $\beta_1$ and $\beta_2$ are dimensionless, $m^*$ and $\Delta m^*_{np}$ are expressed in units of the vacuum nucleon mass $m_N$, while $V_\infty$, $K_0$ and $L$ are expressed in MeV. To obtain the desired momentum dependence of potentials or density dependence of the equation of state, certain parameters of the effective Hamiltonian~\eqref{hamiltonian} are adjusted accordingly.}
\tablab{listfitpar}
\end{table*}

A comparison of transport model predictions for light cluster multiplicities (H and He isotopes) in central AuAu collisions of impact energies in the range 150-800 MeV/nucleon to experimental values~\cite{Reisdorf:2010aa} is presented in ~\figref{multcluster}, using transport model parameters values set to those determined from a fit of experimental data in \secref{fitv1v2}. Results obtained by applying the coalescence algorithm at local freeze-out (bands) and final time of simulations (dashed curves) are shown. For the former case the coalescence parameters have been varied in the ranges $\delta r$=3.0-4.0 fm and $\delta p$=0.2-0.3 GeV/c, while for the latter case the values used in deriving the central result of Ref.~\cite{Cozma:2017bre} have been employed. When the coalescence algorithm is applied at the final time of the reaction typical results emerge: proton multiplicities are overestimated while triton, $^3$He and $^4$He are underestimated. Applying the coalescence algorithm at the local freeze-out time improves the description visibly for all impact energies, model predictions being in satisfactory agreement to experimental data at impact energies equal to or larger than 400 MeV/nucleon. Remaining deviations at larger impact energies, as for example proton multiplicities, can be alleviated by adjusting transport model parameters. A further improvement of the description at the lower end of impact energies (150, 250 MeV) will potentially require explicit cluster degrees of freedom in the transport model~\cite{Danielewicz:1991dh,Wang:2023gta}.

\section{Results}
\seclab{results}
The FOPI collaboration has performed systematic measurements of nucleonic~\cite{Reisdorf:2010aa,FOPI:2011aa} and pionic observables~\cite{Reisdorf:2006ie} in HIC for systems of various sizes and isospin content in the range of impact energies 0.09-1.5 GeV/nucleon. The published database addresses directed and elliptic flows as a function of either reduced rapidity or relative transverse momentum for AuAu collisions~\cite{FOPI:2011aa}. For mid-central collisions of reduced impact parameter 0.25$\leq b_0 \leq$0.45 results for transverse flow of protons, deuterons, $A=3$ clusters and $\alpha$ particles and elliptic flow of protons, deuterons and tritons (and for some impact energies also $\alpha$ clusters) are available. Results for protons in central and mid-peripheral collisions have also been published. Experimental data for an extensive set of stopping observables, in particular $varxz$ for protons, deuterons and tritons, has also been measured~\cite{Reisdorf:2010aa}.


In the present study in-medium cross-sections, momentum dependence of the interaction and the equation of state of nuclear matter are investigated by extracting values for certain model parameters from a comparison of theoretical predictions for $varxz$, transverse and elliptic flows to experimental data in mid-central Ni+Ni, Xe+CsI and Au+Au collisions for five values of the impact energy $T_{lab}$=0.15, 0.25, 0.40, 0.60 and 0.80 GeV/nucleon. We do not compare model predictions to experimental data for the highest three available energies (1.0, 1.2 and 1.5 GeV/nucleon) since during such energetic collisions a non-negligible fraction of nucleons are excited into resonances which may have a non-negligible impact on reaction dynamics at the level of the Vlasov term. The presence of resonance excitation channels still exerts a non-negligible impact, as it will be shown in \secref{fitv1v2}, through the collision term, on the extracted values of in-medium elastic collisions only. Ultimately, this issue can be resolved by including pionic observables in the fit, which will allow fixing the in-medium resonance potentials and implicitly of in-medium modified inelastic cross-sections. Such an extension of the study is postponed for a future work. At the lower end of available energies (0.09 and 0.12 GeV/nucleon), experimental values for elliptic flow become uncertain and their potential to constrain model parameters is thus limited. This is coupled with an increasing importance of Pauli blocking and higher fraction of heavier clusters in the final state as the impact energies is lowered, features that QMD type models have difficulties in reproducing.

\begin{table*}[ht]
\begin{tabular}{|c|l|c|}
\hline\hline
Label & Observables & Reference\\
\hline
FOPI1 & $v_1(y)$, $|y|\leq 0.5$ for Z=1,2 clusters; M5 centrality &A. Andronic et al.\\
&\quad\qquad $^{197}$Au+$^{197}$Au,$^{129}$Xe+CsI, $^{58}$Ni+$^{58}$Ni (0.25, 0.40) & PRC 67, 034907(2003)\\
\hline
FOPI2 & $v_1(y)$, $|y|\leq 0.5$ for p, d, A=3 and $\alpha$ clusters; $0.25 \leq b_0 \leq 0.45$ & W. Reisdorf et al.\\
&\quad\qquad $^{197}$Au+$^{197}$Au (0.25, 0.40, 0.60, 0.80)&NPA 876, 1 (2012)\\
& $v_2(y)$, $|y|\leq 0.5$ for p, d and $\alpha$ clusters; $0.25 \leq b_0 \leq 0.45$&\\
&\quad\qquad $^{197}$Au+$^{197}$Au (0.15, 0.25, 0.40, 0.60, 0.80)&\\
& $v_2(p_T)$, $|y|\leq 0.4$ for p, d and $\alpha$ clusters; $0.25 \leq b_0 \leq 0.45$&\\
&\quad\qquad $^{197}$Au+$^{197}$Au (0.15, 0.25, 0.40, 0.60, 0.80)&\\
\hline
varxz & varxz for p,d,t clusters; $b_0 \leq 0.15$& W. Reisdorf et al. \\
&\quad\qquad  $^{40}$Ca+$^{40}$Ca (0.4, 1.0), $^{58}$Ni+$^{58}$Ni (0.15, 0.25), $^{96}$Ru+$^{96}$Ru (0.4, 1.0) &NPA 848, 366 (2010)\\
&\quad\qquad  $^{96}$Zr+$^{96}$Zr (0.4), $^{129}$Xe+CsI (0.15, 0.25) &\\
&\quad\qquad  $^{197}$Au+$^{197}$Au (0.15, 0.25, 0.4, 0.6, 0.8) &\\
\hline
spectra & longitudinal and transverse rapidity spectra for protons in $Z\leq 3$ clusters&W. Reisdorf et al.\\ 
& \quad\qquad $^{40}$Ca+$^{40}$Ca (0.4), $b_0 \leq 0.15$, $|y_{L,T}/y_P| \leq 1.25$& PRL 92, 232301 (2004)\\
\hline\hline
\end{tabular}
\caption{Experimental data set used to determine the 8 free parameters in \tabref{listfitpar}. Impact energies for each system are given, expressed in GeV/nucleon, in parentheses. Further details regarding cuts in rapidity or transverse momentum for flows and centrality selection for each data set can be found in the corresponding references.}
\tablab{expdataset}
\end{table*}

\subsection{Computational Method}
\seclab{compmethod}

To fulfill the stated goal of the study, simulations with different values of eight of the model parameters have been performed, the complete list and range of variation are listed in~\tabref{listfitpar}. To keep the number of varied parameters to a minimum, the skewness of EoS of symmetric matter and curvature parameter of SE are correlated to the chosen values for $K_0$ and $L$ respectively by the following expressions: $J_0=-600+3.125\cdot(K_0-165)$ [MeV] and $K_{sym}=-488+6.728\cdot L$ [MeV]. These relations correlate, for example, soft (stiff) values for $L$ with negative (positive) values for $K_{sym}$ and are potentially unrealistic since other choices, both at qualitative and quantitative levels, are in principle allowed. Further comments on this issue are provided in~\secref{eos}.

The chosen region in the 8-dimensional parameter space (see~\tabref{listfitpar}) is probed by performing simulations
for 256 parameter sets, selected as uniformly distributed. This is achieved by setting random values to each parameter and ensuring a minimum distance, in parameter space, from previously accepted sets. The minimum distance cut-off is chosen such as to lead to a distribution as close as possible to a uniform one for a reasonable number of randomly generated candidates ($\approx 10^6$). Using theoretical predictions for the 256 parameter sets a model emulator is built for each experimental data point included in the fit. In practice, this is represented by the linear combination of monomials of degree less or equal to $n$=2. Tests with $n$=1 and $n$=3 have also been performed. The former leads to an emulator with a larger model uncertainty, while for the latter over-fitting becomes a problem in certain regions of the probed parameter space. Emulator's robustness has been checked using the leave-one-out cross-validation (LOO-CV) algorithm~\cite{raswill::2006aa}.  

For each choice of parameters the transport model is used to generate 5000 events which ensure a statistical accuracy of theoretical predictions comparable to the experimental one. In addition, the dependence of model predictions on coalescence model parameters is also considered. These are varied independently in the range $\delta r$=[3.0,4.0] fm and $\delta p=$[0.2,0.3] GeV/c, generating 9 values for each observable. The value predicted by the model is determined as the average of these 9 values, while the variance is adopted as systematical uncertainty at 68$\%$ CL. Clearly, the latter choice is rather arbitrary, but is nevertheless fairly conservative. In fact, the so-defined systematic error represents the most important source of uncertainties for the extracted constraints of the eight transport model parameters. In quantitative terms, the systematic model uncertainty is three to four times larger than the statistical one for $v_1$ and two to three times for $v_2$. Probability distributions informing about allowed values for fitted model parameters are determined using the maximum likelihood method~\cite{brandt::2014aa} and employing a Monte-Carlo algorithm that combines importance and stratified sampling~\cite{numrec::1992aa} for an optimal and fast probing of parameter space.

\begin{figure*}[htb]
 \includegraphics[width=0.95\textwidth]{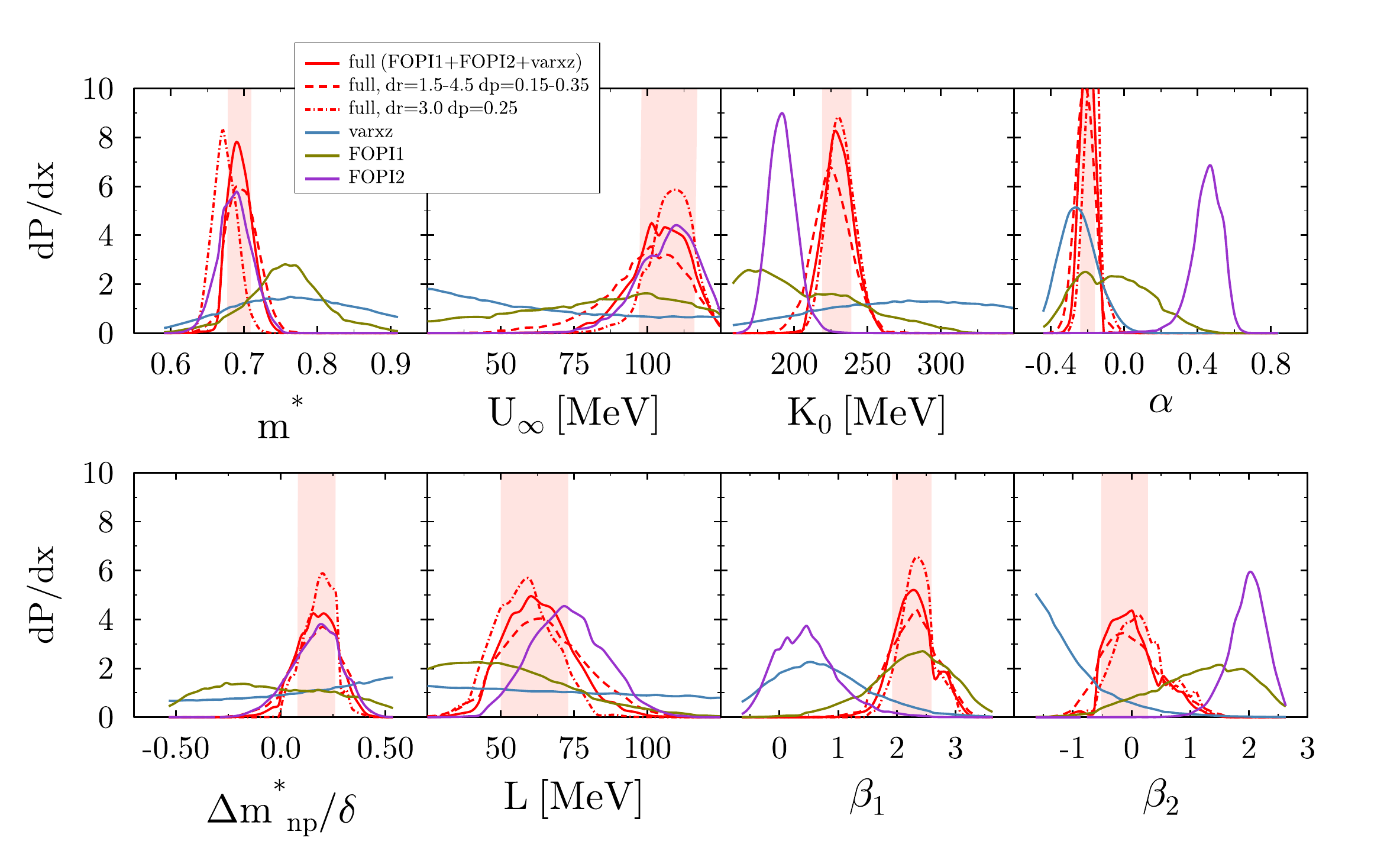}
 \caption{\figlab{fitcombined} (Color online) PDFs of transport model parameter values for the ``full'' case for three different choices of coalescence parameter ranges. Additionally, PDFs for each of the three data sets separately, FOPI1, FOPI2 and varxz, are shown. For each parameter, the 68 $\%$ CL interval is shown by a shaded rectangular area. Average parameter values, 68 $\%$ and 95 $\%$ CL intervals are listed in \tabref{fitflow8par}.}
\end{figure*}

\subsection{Fit to experimental data}
\seclab{fitv1v2}

The experimental data set used in this study to determine the free parameters of the dcQMD transport model is presented in \tabref{expdataset}. It comprises data for rapidity dependent directed flow $v_1(y)$ of $Z=1$, $Z=2$, proton, deuteron, $A=3$ and $\alpha$ fragments at mid-rapidity mainly in Au+Au but also Xe+CsI and NiNi mid-central collisions, rapidity ($v_2(y)$) or transverse momentum ($v_2(p_T)$) dependent elliptic flow for free-protons, deuterons, tritons and $\alpha$ clusters in mid-central Au+Au collisions and the stopping observable $varxz$ for proton, deuteron and triton fragments for 14 systems of various isospin content and impact energy. These three sets of data, labeled FOPI1, FOPI2 and varxz, respectively, make up the ``full'' case mentioned in the following. A fourth set of experimental data, labeled ``spectra'', which only includes longitudinal and transverse rapidity spectra of protons in $Z\leq 3$ fragments in central Ca+Ca collisions at 0.4 GeV/nucleon, was used to asses the robustness of the main result.

Results for transverse momentum dependent directed flow $v_1(p_T)$ are also provided in Refs.~\cite{FOPI:2003fyz,FOPI:2011aa} for various ranges of scaled rapidity within the interval $0.4 \leq y/y_P \leq 1.1$. It has proven however impossible to describe proton $v_1(p_T)$ for values of the scaled transverse momentum $p_T/p_P\leq 1.0$. In this region the model significantly overshoots the experimental data, suggesting contamination by deuterons and tritons. A similar effect, but not as severe, is observed for $Z=1$ (but not $Z=2$) clusters which suggests that proton $v_1(p_T)$ is contaminated also by helium and possibly heavier fragments. Additionally, in this low scaled transverse momentum region the systematic uncertainty due to variation of coalescence model parameters is exceptionally high. A similar effect is seen also for $v_1(y)$ for rapidities away from mid-rapidity for $T_{lab}>0.2$ GeV/nucleon and for almost the entire rapidity range for Au+Au collisions at 0.15 GeV/nucleon. This type of problem is not completely unexpected. In the mentioned range of rapidities particles originating from the fragmentation process of spectators are encountered and shortcomings of the simple coalescence model employed become apparent. Consequently, available observables for $|y/y_P|>0.5$ have not been included in the set of fitted data. Including them will result in a much softer compressibility modulus, below the lower limit, $K_0=165$ MeV, simulated in the study. Comparison of predictions for $v_1(p_T)$ to selected experimental data are nevertheless presented in Appendix A.

\begin{table*}[ht]
\begin{tabular}{|c|c|c|c|c|c|c|}
\hline\hline
\multicolumn{1}{|c|}{} &\multicolumn{3}{|c|}{Full} & \multicolumn{3}{|c|}{Full, No Threshold Effects}\\
\hline
Parameter & Average & 68\% CL & 95\% CL & Average & 68\% CL & 95\% CL\\
\hline
$m^{*}$ & 0.695& [0.677,0.709]&[0.669,0.725] & 0.684& [0.669,0.696]&[0.654,0.718]\\
$U_\infty$ & 104& [97,116]&[86,124] & 116& [111,124]&[102,127]\\
$K_0$ & 230& [219,239]&[210,250]& 228& [218,238]&[208,246]\\
$\alpha$ & -0.20& [-0.24,-0.16]&[-0.27,-0.13]& -0.10& [-0.15,-0.06]&[-0.20,-0.02]\\
$\Delta m^*_{np}/\delta$ & 0.17& [0.08,0.27]&[0.0,0.36]& -0.46& [-0.54,-0.44]&[-0.54,-0.36]\\
$L$ & 63& [50,73]&[41,85]& 118& [107,134]&[91,142]\\
$\beta_1$ & 2.28& [1.92,2.59]& [1.60,3.01]& 2.77& [2.50,3.03]&[2.18,3.25]\\
$\beta_2$ & 0.03&[-0.52,0.28] &[-0.60,1.12]& 0.81& [0.50,1.21]&[0.08,1.48]\\
\hline\hline
\end{tabular}
\caption{Result of the fitting procedure that includes the first three sets of experimental data in \tabref{expdataset} for two cases: the full model and full model with threshold effects switched off. Average values for each of the 8 varied parameters and their 68$\%$ and 95$\%$ confidence level intervals are listed. Parameters $U_\infty$, $K_0$ and $L$ are expressed in units of MeV, while the rest are dimensionless.}
\tablab{fitflow8par}
\end{table*}

The result of fitting the chosen experimental data sets is shown in \figref{fitcombined} as probability distribution functions (PDFs) for each parameter. For the ``full'' case, PDFs for three choices of the coalescence parameter ranges are provided: the standard choice ($\delta r$=3.0-4.0 fm, $\delta p$=0.2-0.3 GeV/c), a maximum ranges choice ($\delta r$=2.5-4.5 fm, $\delta p$=0.15-0.35 GeV/c) and a choice consisting of definite values for the two coalescence parameters ($\delta r$=3.0 fm, $\delta p$=0.25 GeV/c). The first choice corresponds to a range that best describe multiplicities of clusters with $Z \leq 2$ in central Au+Au collisions, see \figref{multcluster}. The second one includes suboptimal choices, while the third corresponds to a close to optimal choice. The variation in the location of the maximum of PDFs quantifies the potential residual model dependence due to coalescence. It is seen to always amount to less than one standard deviation.

\begin{figure}
\includegraphics[width=0.495\textwidth]{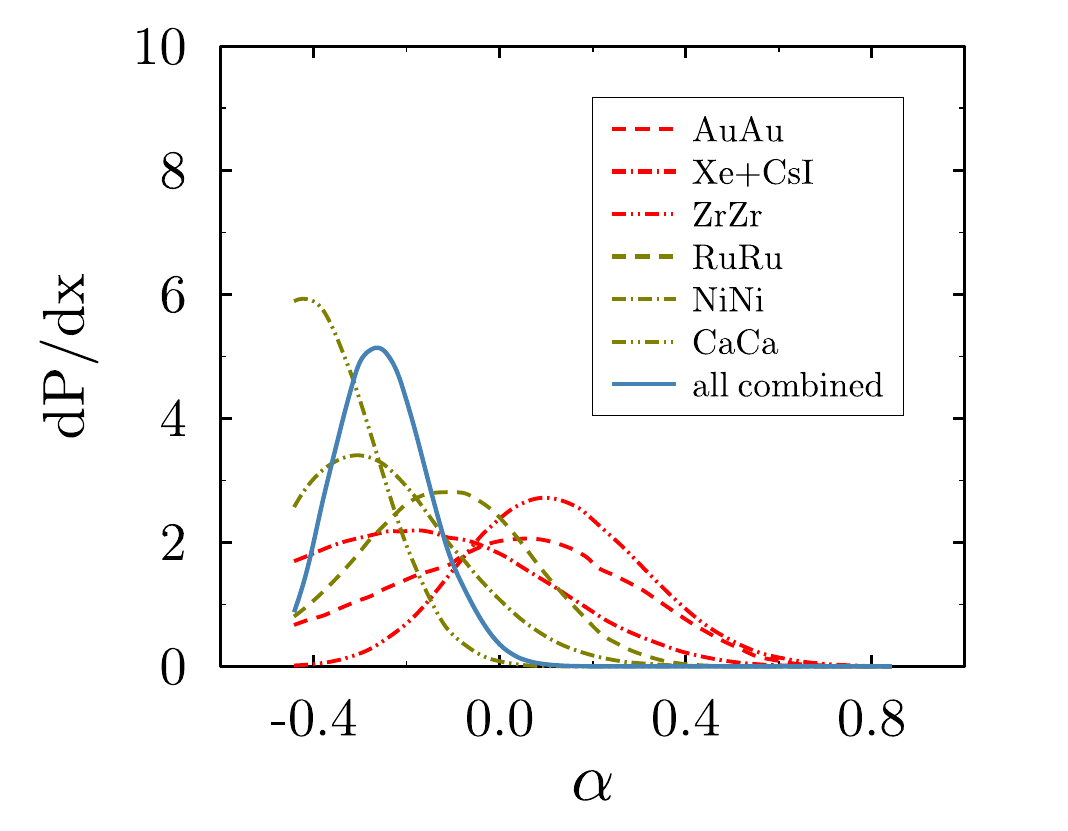}
\caption{\figlab{stopping} (Color online) PDFs of $\alpha$ derived from fitting $varxz$ data for each cluster species separately. Combined result for all 14 systems is shown by the full curve.}
\end{figure}

\begin{figure*}
 \includegraphics[width=1.0\textwidth]{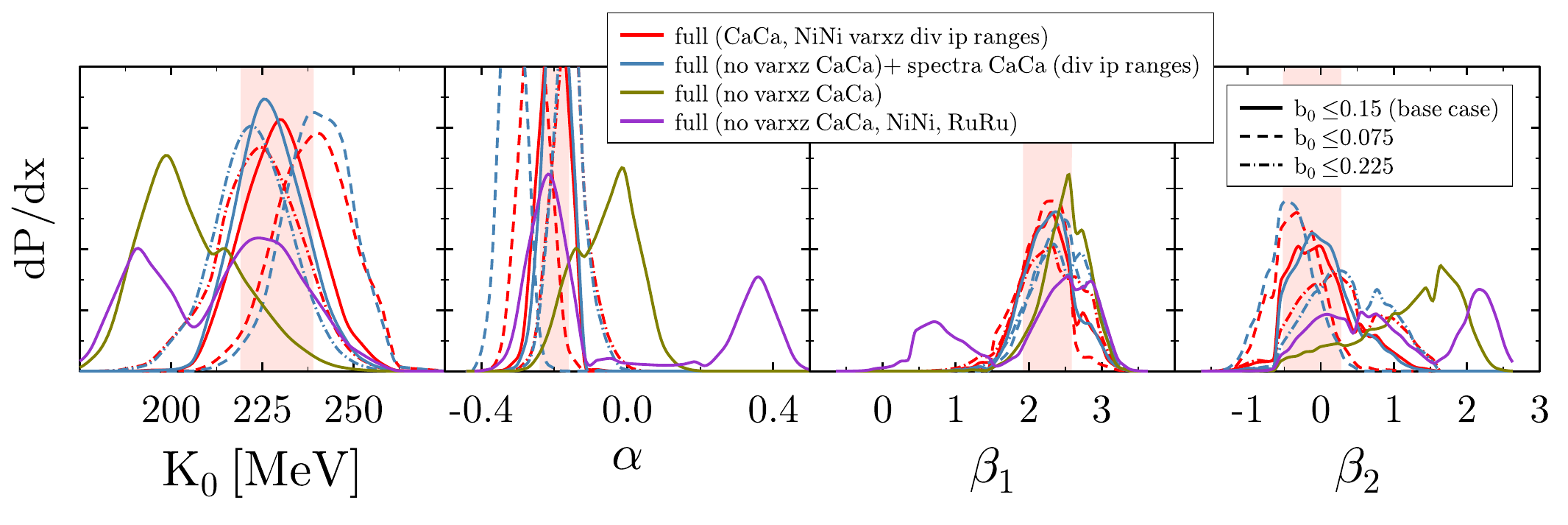}
 \caption{\figlab{fitcaca} (Color online) Sensitivity of the compressibility modulus $K_0$ to $varxz$ for isospin symmetric systems and uncertainties in the impact parameter determination for light systems. Some impact on the in-medium modification of cross-sections in isospin asymmetric matter is also observed. Curve color codes the list of data included in the fit (variations of the ``full'' case by removing or replacing varxz for CaCa at 0.4 GeV/nucleon with proton longitudinal and transverse rapidity spectra for $Z \leq 3$ for the same reaction; a case with the varxz data for all (nearly) isospin symmetric systems is also shown). Curve styles denotes different reduced impact parameter ranges for simulated CaCa and/or NiNi systems. Shaded areas correspond to the 68$\%$ CL level ranges of fitted parameters for the ``full'' case.}
\end{figure*}

Additionally, PDFs for model parameters for each of the first three data sets in ~\tabref{expdataset}, fitted individually, are also shown. The stopping observable $varxz$ displays significant sensitivity only to parameters that adjust the in-medium correction factors of elastic cross-sections: $\alpha$, $\beta_1$ and $\beta_2$. A more detailed investigation, not shown here, reveals that $varxz$ imposes a correlation between $\alpha$ and the iso-scalar effective mass parameter $m^{*}$. The experimental data set FOPI1, consisting of directed flow data $v_1(y)$ at mid-rapidity, exhibits noteworthy sensitivity to in-medium cross-sections and isoscalar effective mass. The FOPI2 data sets presents sensitivity to all parameters, which is to a large extent due to the transverse momentum dependent elliptic flow $v_2(p_T)$. The difference in the favored values for in-medium cross-section between these two sets of data is the result of including $v_1(y)$ observable for NiNi collisions in the former, which underlines the importance of studying systems with different average isospin asymmetry. 

By comparing ``full'' and FOPI2 cases a correlation between $K_0$ and $\alpha+\beta_1$ is noticed. The large difference between the preferred values for $\alpha$ for the two cases is due to the stronger in-medium suppression of cross-sections for light isospin symmetric systems required to describe $varxz$ observables, in particular for CaCa at 0.4 GeV/nucleon, see \figref{stopping}. It is seen that, on average, CaCa favors $\alpha\approx -0.4$ while for AuAu $\alpha\approx 0.1$. The other two parameters fixing in-medium cross-sections, $\beta_1$ and $\beta_2$ display almost flat distributions for separate systems (not shown). As a result, in-medium cross-sections in CaCa HIC are suppressed by an additional factor 0.6 as compared to AuAu. There are several possible explanations for this apparent strong effect.

A first possibility is related to a potentially less efficient Pauli blocking algorithm in lighter systems due to stronger fluctuations induced by lower total number of particles. Tests reveal that this is in fact the case, however Pauli blocking in Ca+Ca system is less effective by only about 10$\%$ as compared to AuAu. A second possibility is a more subtle systematic uncertainty induced by the coalescence model, since used $varxz$ data are only available for protons, deuterons and tritons. To test this hypothesis, $varxz$ data for CaCa at 0.4 GeV/nucleon are replaced by longitudinal and transverse proton rapidity spectra for fragments with $Z\leq 3$~\cite{FOPI:2004orn}, which are close to coalescence invariance. The results, see \figref{fitcaca}, show an impact on both $\alpha$ and $K_0$ well within the 68$\%$ CL interval. Removing CaCa $varxz$ (or rapidity spectra) from the fit leads to lower value for $K_0$ (the average value $K_0$=205 MeV is slightly outside the $95\%$ CL interval for the full case) and milder suppression of cross-sections $\alpha=-0.14$. Removing additionally NiNi and RuRu $varxz$ has stronger impact on final values, the PDFs for both $\alpha$ and $K_0$ becoming bimodal, which can be traced back to the values favored by the FOPI1 and FOPI2 data sets if fitted separately, see \figref{fitcombined}. Important impact is observed also on the preferred values for $\beta_1$ and $\beta_2$.

At third source of possible uncertainties is the impact parameter range. Due to finite number fluctuations, impact parameter determination for light systems is uncertain~\cite{FOPI:2006ifg}. To investigate this possibility the impact parameter range for simulations of central CaCa and NiNi collisions is changed from the standard one, $b_0\leq 0.15$, to either $b_0\leq 0.075$ or $b_0\leq 0.225$. The impact on both $K_0$ and $\alpha$ is not negligible, particularly for the latter choice, but is confined to the 68$\%$ CL interval. This is the result of a strong dependence of $varxz$ on impact parameter~\cite{FOPI:2010xrt}. Additionally, the impact of the rapidity cut on rapidity spectra has been checked. Including only the region of spectra with $|y_{L,T}|/y_P \leq 0.75$ lead to a constraint for $K_0$ and $\alpha$ similar to the case ``full (no varxz CaCa)'' in \figref{fitcaca}. This is understandable, since $varxz$ is determined mainly by the region of rapidity spectra $0.75\leq |y_{L,T}|/y_P \leq 1.25$. It is concluded that the``full'' case result is robust, however potential systematical uncertainties of magnitude comparable to the 68$\%$ CL interval due to less effective Pauli blocking algorithm and uncertainties in the experimental determination of the impact parameter for light systems cannot be excluded.

The relevance of several model ingredients is studied by determining their impact on PDFs, see \figref{fitcombined2}. In previous studies~\cite{Cozma:2017bre} the coalescence model has been applied at the end to the evolution of the system, typically at t=150 fm/c at which moment the system had reached its asymptotic state. As shown in \secref{coalescence} this has strong impact on cluster multiplicities which is reflected strongly on the favored magnitude of in-medium cross-sections and momentum dependence of the interaction in both isospin SNM and ANM. The impact on the compressibility modulus is large, the average value $K_0$=175 MeV is close to the lower limit of the 99.8$\%$ CL interval of the full result. The impact on the value of $L$ is surprisingly small.

\begin{figure*}[htb]
 \includegraphics[width=0.95\textwidth]{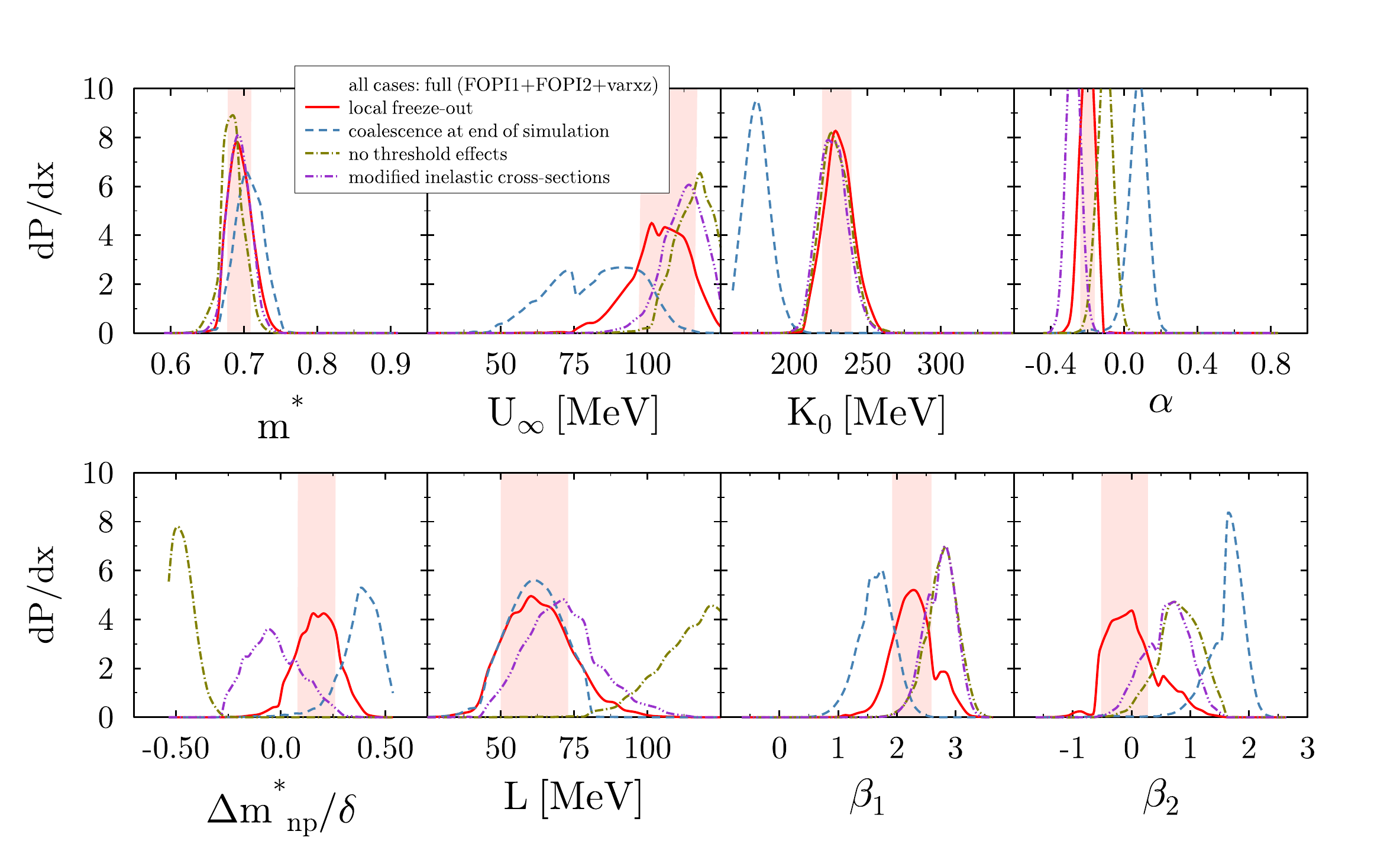}
 \caption{\figlab{fitcombined2} (Color online) PDFs of model parameter values for the ``full'' case and three additional scenarios: coalescence applied at the end of simulations, threshold effects switched off and modification of inelastic $NN$ cross-sections. For each parameter, the 68 $\%$ confidence level interval is shown by a shaded area. Average parameter values, 68 $\%$ and 95 $\%$ CL intervals are listed in \tabref{fitflow8par}.}
\end{figure*}

\begin{figure*}[htb]
 \includegraphics[width=0.95\textwidth]{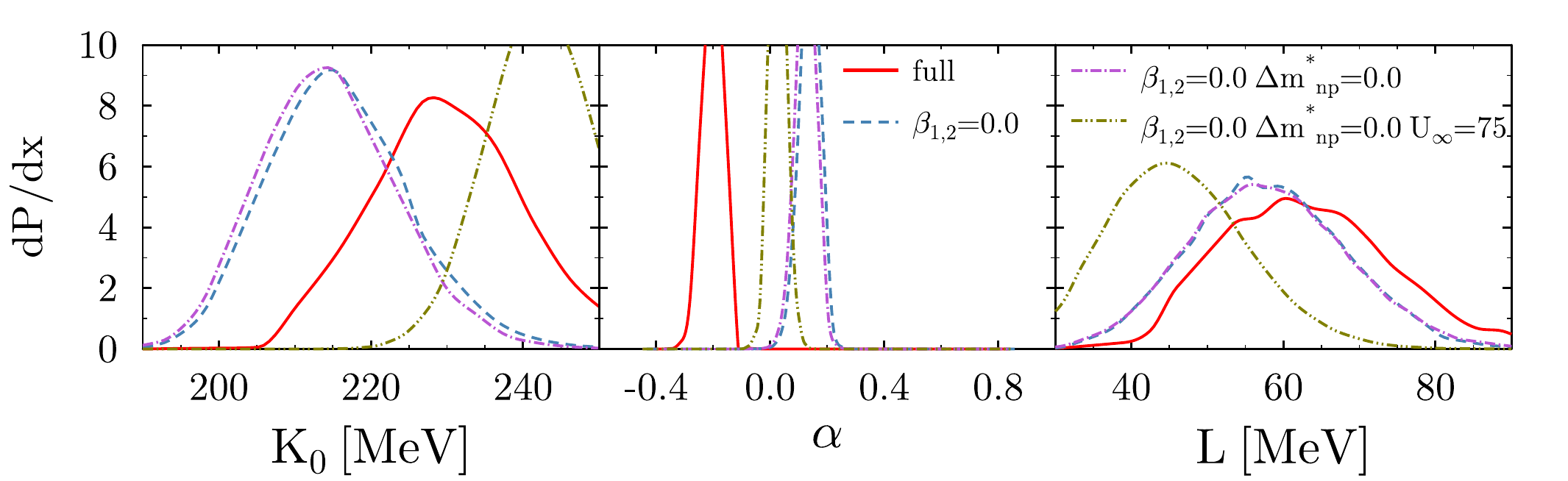}
 \caption{\figlab{fitcombined3} (Color online) Impact of setting $\beta_1=0.0$ and $\beta_2=0.0$, $\Delta m^*_{np}$=0.0 and $U_\infty$=75 MeV sequentially on EoS parameters $K_0$ and $L$ and in-medium modification factor of elastic $NN$ cross-section in SNM $\alpha$.}
\end{figure*}

Another important model ingredient is the inclusion of threshold shifts that arise in dense matter as results of the momentum dependence of the interaction. They lead to strong effects for particle production close to threshold and are crucial for the description of pion production in HIC and study of the SE using pionic observables~\cite{Ferini:2006je,Cozma:2014yna,Song:2015hua,Ikeno:2023cyc}. Their impact for elastic $NN$ collisions is expected to be smaller and previous estimates support this idea. Due to the particular momentum dependence of the interaction (more attractive at lower momenta in rest frame of NM) elastic collisions for particles with the same kinetic momentum as in vacuum are above the in-medium threshold by a larger amount of energy than in vacuum. Since elastic vacuum $NN$ cross-sections decrease with increasing invariant mass, one expects that the effect of threshold effect on in-medium $NN$ cross-section is to decrease them. This effect is compensated by a larger value of $\alpha$ during the calibration of the model to experimental data. This expectation is confirmed for both the isoscalar and isovector channels, see \figref{fitcombined2}. The impact on the momentum dependence of the interaction result in both the isoscalar optical and isovector Lane potential becoming more repulsive at large momenta. The impact on $\Delta m^*_{np}$ is particularly important, omission of threshold effects leading to a negative value for this parameter, in conflict to most studies in the literature. The favored average value of the slope parameter $L=118$ MeV and its PDF depart from the full result at more than $95\%$ CL. Numerical results for average values of all 8 parameters, as well as their $68\%$ and $95\%$ CL intervals for this case are listed in \tabref{fitflow8par}.

Lastly, the impact of in-medium modifications of inelastic $NN$ cross-sections has been estimated. For simplicity, values of effective masses entering in the expression of cross-section (incoming flux and final-state phase-space factor) are set to vacuum ones. This results in an increase of cross-sections by a factor that is most often below 2. This effect is expected to be compensated by a lower value of $\alpha$ in the isoscalar, which is indeed observed, see \figref{fitcombined2}. In the isovector channel the opposite holds. The impact on the density dependence of EoS of SNM and ANM is found to be small. The momentum dependence of the isoscalar and isovector potentials is modified to a slightly more repulsive one at high momenta, see \figref{fitcombined2}.

The necessity of fitting parameters $\beta_1$, $\beta_2$, $\Delta m^*_{np}$ and $U_\infty$ is addressed in \figref{fitcombined3}. They are set, in sequence, either to zero (the first three) or to their usual values ($U_\infty$= 75 MeV~\cite{Cozma:2017bre}). Isospin dependence and splitting of transition amplitudes are switched off simultaneously due to a strong correlations between values of $\beta_1$ and $\beta_2$. The impact on $K_0$ and $\alpha$ amounts to approximately two standard deviations. The slope of SE $L$ is significantly less affected. Setting $\Delta m^*_{np}$=0.0 has little effect on parameters $\alpha$, $K_0$ and $L$, its impact on the isoscalar effective mass $m^*$ is however close to one standard deviation (not shown). Lastly, setting $U_\infty$=75 MeV affects all three parameters, particularly $K_0$ (three standard deviations), but also $L$ (two standard deviations). This last point underlines the importance of momentum dependence of interactions for EoS studies and the need to employ one that is compatible with heavy-ion data over the entire range of probed momenta.

\section{Discussion}
\seclab{discussion}
The implications of the results of previous section are presented by determining the sensitivity of flow observables to the EoS of SNM and ANM followed by a presententation, in comparison with available results in the literature, of the contraints for EoS, in-medium elastic cross-sections and momentum dependence of the optical potential deducted in this study.

\subsection{Probed densities}
\seclab{probeddens}

The process of heavy-ion collisions consists of two main stages: compression, while at the core of the reaction region a fireball with peak density in the 1.5-3.0$\rho_0$ range is formed, depending on impact energy, followed by an expansion phase during which the system attains its final state comprised of free-nucleons and clusters and when densities below saturation are also probed. Final values of observables of interest are the result of accumulated effects during the entire history of the process. It is of interest to know which range in density has the largest impact and what is the average, with respect to sensitivity, probed density. To this end, it is important to recognize that observables are functional of the EoS, (in-medium) cross-sections, optical potentials and other quantities. The sensitivity with respect to the EoS at a particular density $\tilde\rho$ can be defined as the functional derivative of the observable of interest with respect to the EoS
\begin{eqnarray}
\eqlab{sens1}
 \frac{d \mathcal{O}}{d E/N}(\tilde\rho)&=&\lim_{\varepsilon \to 0} \frac{1}{2\varepsilon}\,\bigg[\mathcal{O}\Big(E/N(\rho)+\varepsilon \delta(\rho-\tilde\rho)\Big)\\
 &\qquad&\qquad\qquad-\mathcal{O}\Big(E/N(\rho)-\varepsilon \delta(\rho-\tilde\rho)\Big)\bigg]\nonumber
\end{eqnarray}
or
\begin{eqnarray}
\eqlab{sens2}
 \frac{d \mathcal{O}}{d (E/N)'}(\tilde\rho)&=&\lim_{\varepsilon \to 0} \frac{1}{2\varepsilon}\,\bigg[\mathcal{O}\Big(\frac{d E/N}{d \rho}(\rho)+\varepsilon \delta(\rho-\tilde\rho)\Big)\\
 &\qquad&\qquad\qquad-\mathcal{O}\Big(\frac{d E/N}{d \rho}(\rho)-\varepsilon \delta(\rho-\tilde\rho)\Big)\bigg].\nonumber
\end{eqnarray}

\begin{figure}
\includegraphics[width=0.495\textwidth]{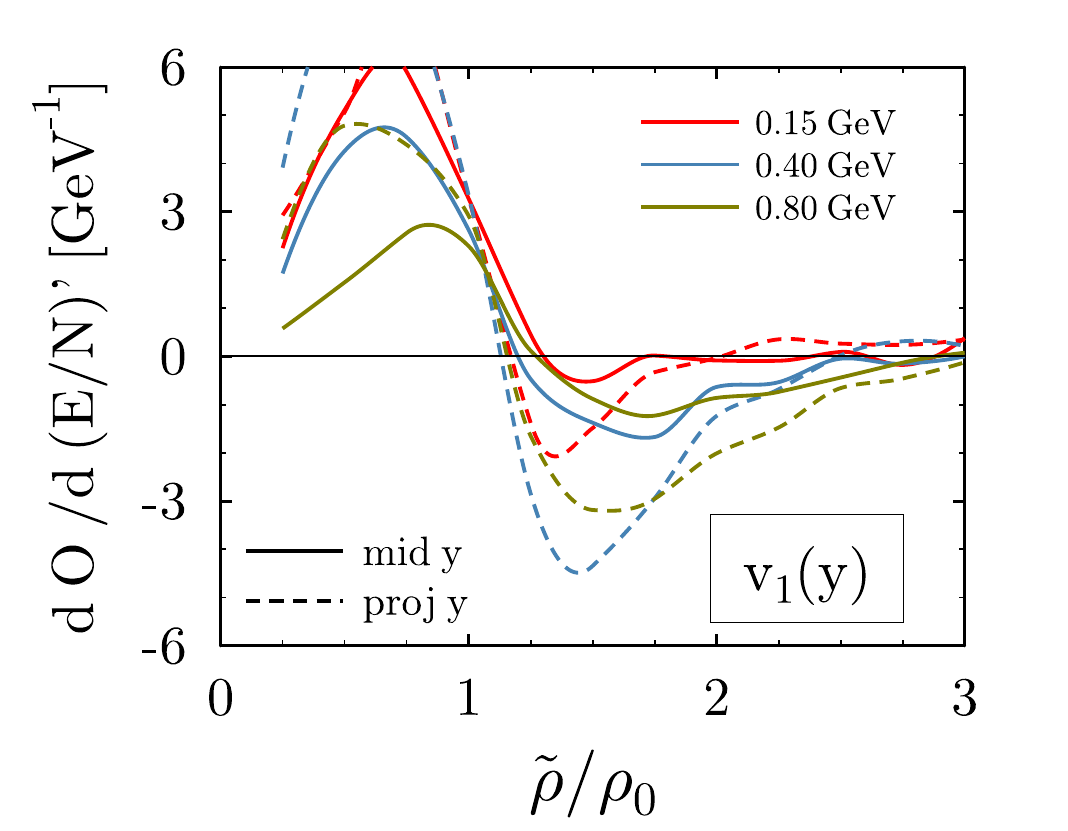}
\caption{\figlab{denssens_k0prot} (Color online) Sensitivity of rapidity dependent directed flow $v_1(y)$ of protons to the density dependence of SNM as function of density for mid-central ($0.25<b_0<0.45$) AuAu collisions. Color codes impact-energy of the projectile, while line style denotes range of rapidity: mid-rapidity (``mid y'') and projectile rapidity (``proj y'') correspond to $0.0 \leq y/y_P \leq 0.5$ and $0.5\leq y/y_P \leq 1.0$ respectively. For all cases a cut in transverse momentum was also applied: $p_T/p_P \geq 0.4$.}
\end{figure}

\begin{figure*}
\begin{center}
 \includegraphics[width=0.9\textwidth]{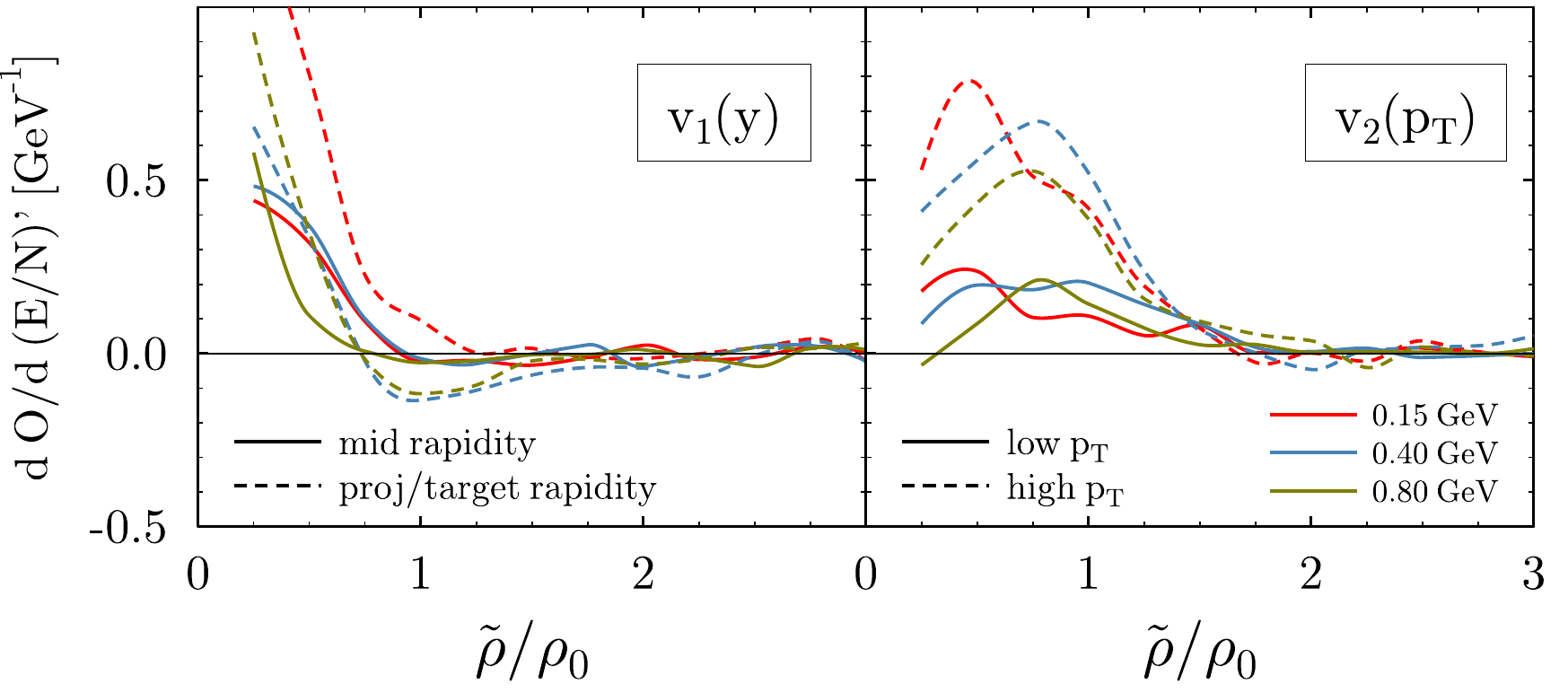}
 \end{center}
 \caption{\figlab{denssens_lsymprot} (Color online) Sensitivity of rapidity dependent directed flow $v_1(y)$ (left panel) and transverse momentum dependent elliptic flow $v_2(p_T)$ (right panel) of protons to the density dependence of symmetry energy as function of density for mid-central ($0.25<b_0<0.45$) AuAu collisions. In both panels color denotes impact energy of the projectile. The same details as in \figref{denssens_k0prot} are in order for $v_1(y)$ (left panel). For $v_2(p_T)$ (right panel) ``low $p_T$'' and ``high $p_T$'' correspond to $0.4 \leq p_T/p_P \leq 1.2$ and $1.2 \leq p_T/p_P \leq 2.0$ respectively. A cut in rapidity, $0.0 \leq y/y_P\leq 0.4$, that mirrors the experimental filter was also applied in this case.}
\end{figure*}

The two expressions above are equivalent since knowing the derivative of a function is equivalent to knowing the function up to an arbitrary constant. In the case of the EoS, we have fixed the binding energy at saturation density for SNM and the magnitude of SE at $\rho$=0.1 fm$^{-3}$. The former is disadvantageous, from a computational point of view, when the main contribution to sensitivity originates from the mean-field propagation through terms proportional to the derivative of the energy per nucleon with respect to density, as is the case for the sensitivity to the EoS of SNM and that of SE proportional with the gradient of local density (drift term). It can however be used when threshold effects have a strong impact on observables ($e.g.$ particle production close to threshold) or for the term in the mean-field contribution involving the SE and the gradient of the local isospin asymmetry (diffusion term). For the purpose of this study, the latter definition is more useful, as it induces a local modification of either $L_0$ (slope of the EoS of SNM) or $L$ in the vicinity of target density $\tilde\rho$.

In practice, the Dirac $\delta$ function appearing in \eqref{sens1} and \eqref{sens2} is replaced by a Gaussian of width $\eta$. To account for the modification of the interaction, the following terms are added to the expression of the Hamiltonian of the system in \eqref{hamiltonian}
\begin{eqnarray}
\Delta \langle H_{L_0} \rangle &=&\sum_{i=n,p} \frac{\varepsilon}{2}\,\bigg [Erf\Big(\frac{\tilde u}{\eta}\Big)-Erf\Big(\frac{\tilde u-u_i}{\eta}\Big)\bigg ]  \\
\Delta \langle H_{L} \rangle &=&\sum_{i=n,p} \frac{\varepsilon}{2}\,\tilde\tau_i\,\delta_i\,\bigg[Erf\Big(\frac{\tilde u}{\eta}\Big)-Erf\Big(\frac{\tilde u-u_i}{\eta}\Big)\bigg] \nonumber
\end{eqnarray}
for SNM and SE respectively. The two additional parameters $\eta$ and $\varepsilon$ take the following values: $\eta$=0.125 while $\varepsilon$=0.01 GeV and $\varepsilon$=0.05 GeV for SNM and SE sensitivity calculations respectively. It should be noted that sums in each correction term run only over nucleons. Also, the reduced density $u_i$ should only include contributions from nucleons in case $NN$ and $N\Delta$ interactions are different (see \cite{Cozma:2021tfu}). Clearly, as more nucleons are excited to resonances, the sensitivity to $\it pure\,nucleonic$ EoS is reduced. 

We start by reporting the sensitivity to $L_0(\tilde\rho)$ of directed flow of protons, see \figref{denssens_k0prot}. Calculations for AuAu collisions at three impact energies 0.15, 0.4 and 0.8 Gev/nucleon have been performed. The values listed in \tabref{fitflow8par} for the eight free model parameters for the full case, including threshold effects, have been used. The rapidity interval $0.0<y/y_P<1.0$ has been split into two regions for which results are presented separately. The dominant contribution to sensitivity is found to originate at sub-saturation densities. However, with increasing impact energy, a larger fraction of sensitivity is found at densities above $\rho_0$. At 0.8 GeV/nucleon there is comparable sensitivity to the density intervals [1.0,2.0]$\rho_0$ and [2.0,3.0]$\rho_0$. Clearly, if the EoS of SNM is accurately known below $\rho_0$ then the EoS around 2$\rho_0$ and above can be extracted from HICs of increasing impact energy. At densities close to 1.25$\rho_0$ the sensitivity vanishes for all cases, which is the consequence of the existence of the saturation point where the isoscalar component of the force vanishes. The crossing density is larger than $\rho_0$ as a result of the migration of the saturation density towards higher values as temperature increases. Similar results were obtained for other flow observables, $v_1(p_T)$, $v_2(y)$ and $v_2(p_T)$. Additionally, other particle species, in particular $n$, $d$, $t$ or $A=3$ and $\alpha$, display similar sensitivity to $K_0(\tilde\rho)$. It should be stressed again that the expression in \eqref{sens2} is a functional of EoS, cross-sections and other quantities. For example, changing in-medium elastic cross-sections to be closer to vacuum ones, by setting $\alpha$=0.2, the sensitivity of $v_1(y)$ to suprasaturation densities for projectile rapidity particles is strongly reduced, however it survives at mid-rapidity.

A similar calculation for $L(\tilde\rho)$ is reported in \figref{denssens_lsymprot}. A somewhat different behavior is found for directed and elliptic flow of protons, hence results for both $v_1(y)$ and $v_2(p_T)$ are shown. The former probes only sub-saturation densities, while the latter shows sensitivity up to about 1.5$\rho_0$. The same result, not shown here, was obtained for other $Z\geq 1$ clusters. The situation appears more promising for neutrons: mid-rapidity $v_1(y)$ probes densities up to 1.5$\rho_0$, while $v_2(p_T)$ for large transverse momenta probes densities up to twice saturation. As before, simulations at three different impact energies were performed: 0.15, 0.40 and 0.80 GeV/nucleon. Surprisingly, no sizable dependence on the impact energy was found for any clusters, including neutrons. 

\begin{figure*}
\includegraphics[width=0.75\textwidth]{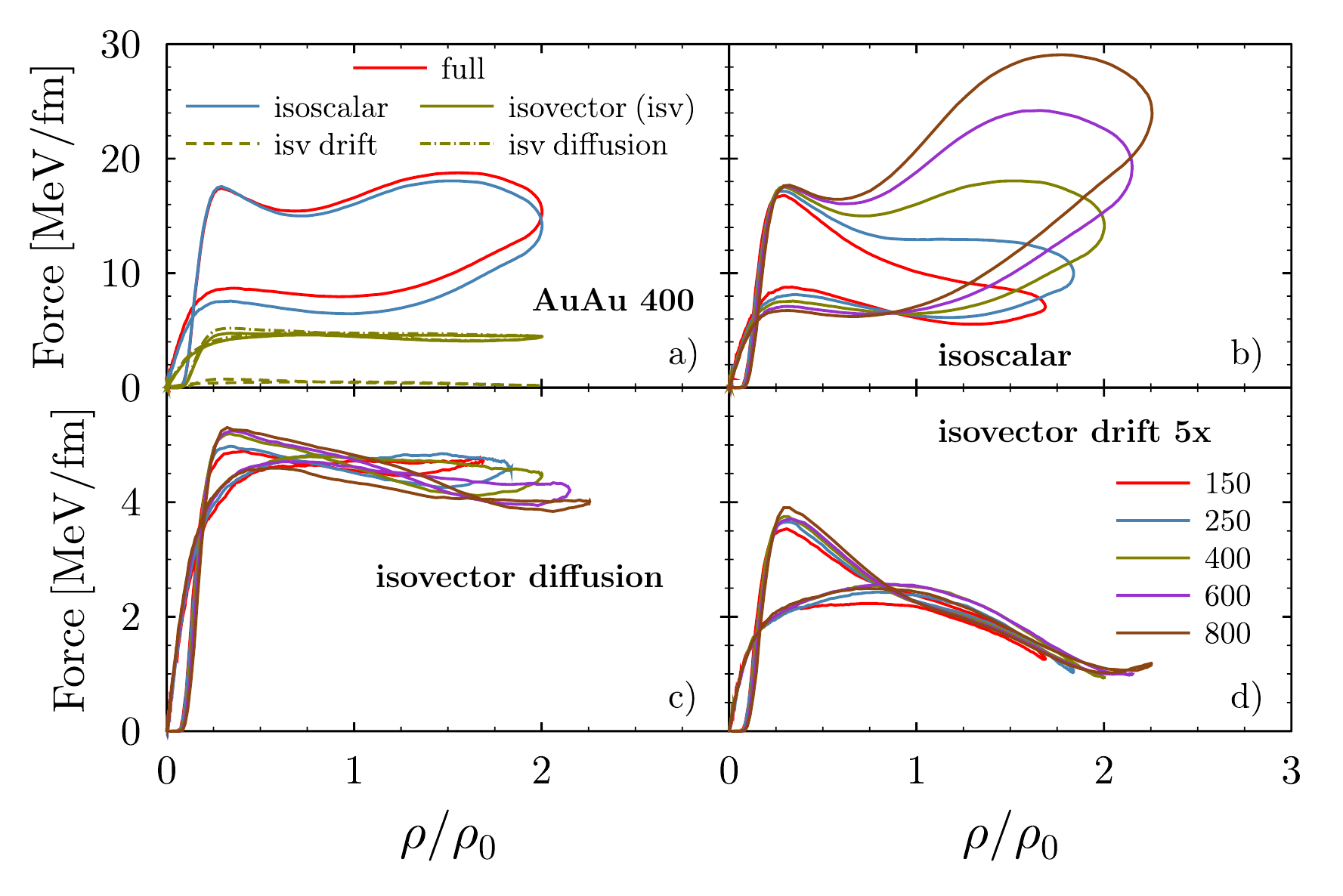}
\caption{\figlab{force} (Color online) Average trajectories of HIC in the force-density plane for matter in a sphere of radius R=3.0 fm located at the origin of the coordinate systems. The force is determined as average, over nucleons, of the modulus of force in the Vlasov term. Panel a) shows trajectories of the total (including Coulomb), strong isoscalar and isovector for mid-central AuAu collisions. The two components of the isovector force, the drift and diffusion terms, are also shown explicitly. Panels b), c) and d) present trajectories of the isoscalar, diffusion and drift components for mid-central AuAu collisions at five beam energies.}
\end{figure*}

A proper understanding of the difference between sensitivity for SNM and ANM can be achieved by determining the average, over nucleons, of the magnitude of force in the Vlasov term of the equations of motion in a sphere of radius $R=$ 3 fm located at the origin of the coordinate system, see \figref{force}. The evolution of the system is presented as trajectories in the force-density plane in each panels. In panel a) the total force (including Coulomb) and its strong isoscalar and isovector components are shown for mid-central Au+Au collisions at 400 MeV/nucleon beam energy. The two components of the isovector force, the drift and diffusion terms are also shown explicitly, the former being negligible in relative terms. The isoscalar component provides the dominant contribution to the total force.

Panels b), c) and d) of \figref{force} present trajectories in the force-density plane for each of the three components of the strong force (isoscalar, diffusion and drift) in mid-central Au+Au collisions for five beam energies: 150, 250, 400, 600 and 800 MeV/nucleon. A clear difference is observed between the isoscalar and isovector components. The former shows a strong dependence on beam energy: at maximum reached density the isoscalar force at 800 MeV/nucleon is about three times stronger than at 150 MeV/nucleon. In contrast, the isovector diffusion term becomes moderately weaker with increasing beam energy, while the drift term stay approximately the same. These trends are related to similar trajectories in the $\nabla \rho$-$\rho$, $\nabla \delta- \rho$ and $\delta\nabla \rho - \rho$ planes respectively. 

The observed behavior of the isoscalar and isovector components of the force appears in one-to-one correspondence with the beam energy dependence of sensitivity of flow observables with respect to EoS of SNM and ANM respectively. Loosely speaking, the latter is obtained by integrating the force over the duration of the evolution. The entire process of compression-expansion during HIC requires less time with increasing impact energy. Consequently, the strong dependence of isoscalar effects on beam energy, shown in panel b) of \figref{force} is expected to become quenched. An analysis, with similar results, of the collision term, which exhibits a dependence on the interaction through threshold effects, has also been performed. Changing the density dependence of the SE to a stiff one, $L$=123 MeV and $K_{sym}$=340 MeV leads to stronger isovector forces, increasing with density. The drift component of the force now exhibits an important dependence on beam energy (however the diffusion term does not), yet flow sensitivities to SE do not mirror this behavior.


\begin{figure*}[htb]
\includegraphics[width=0.495\textwidth]{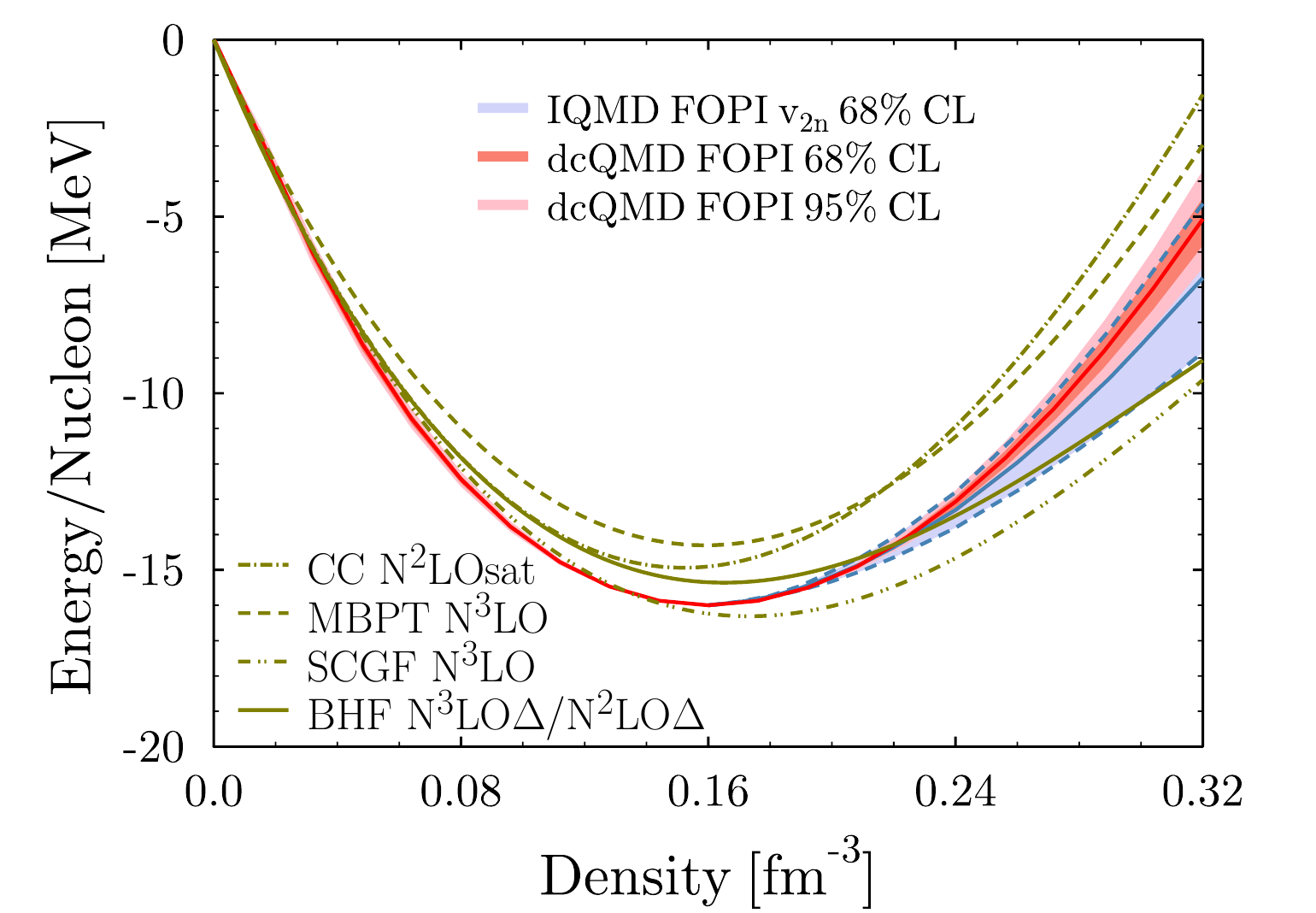}
\includegraphics[width=0.495\textwidth]{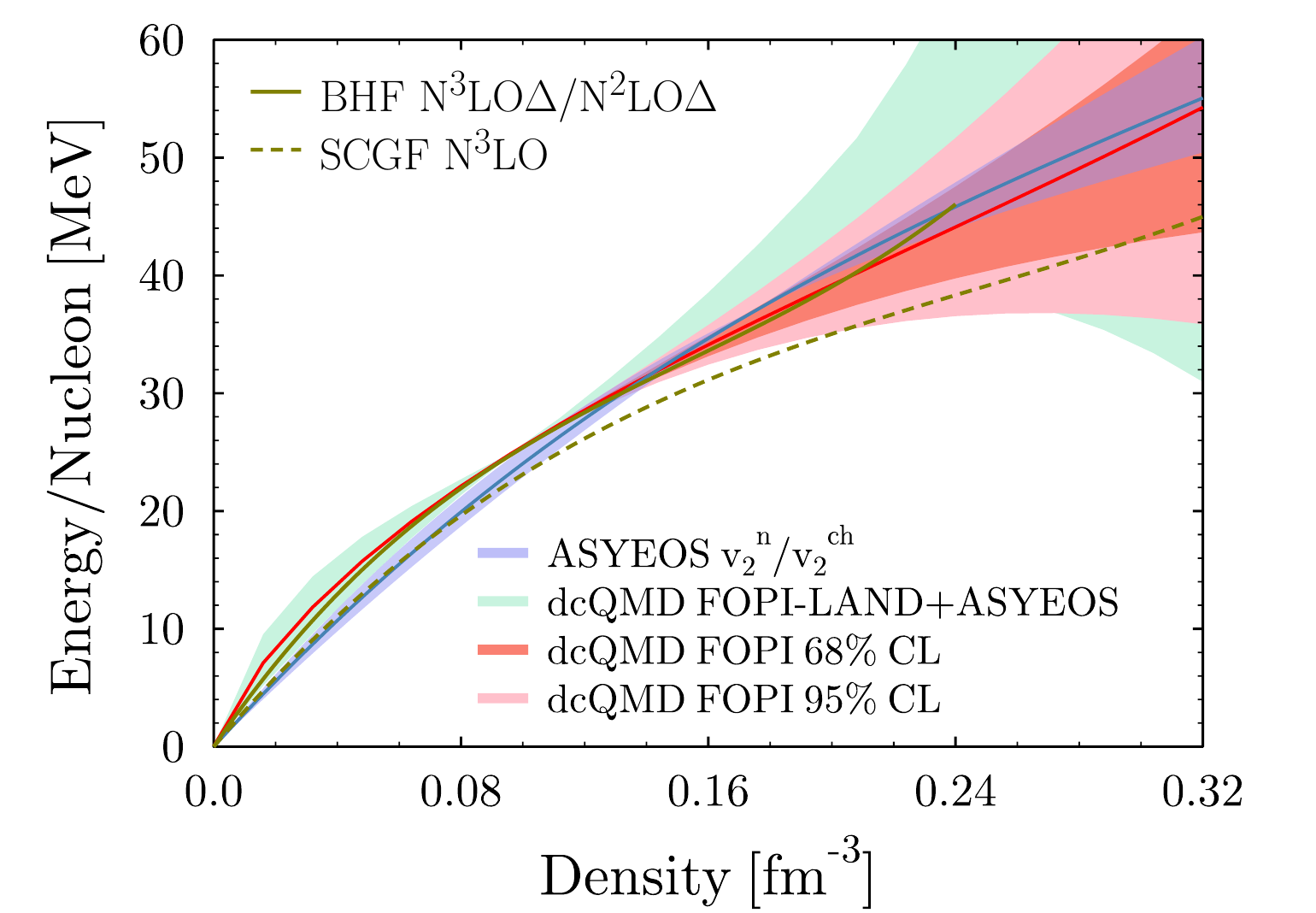}
\caption{\figlab{eos_snm_se} (Color online) Comparison of the constraints obtained in this work for the density dependence of EoS of SNM (left panel) and SE (right panel) to other analysis of HIC data and ab-initio calculations, see text for a detailed explanation. }
\end{figure*}

\subsection{Equation of state}
\seclab{eos}
One of the central results of the study are values for the compressibility modulus $K_0$ of SNM and the slope parameter $L$ of SE at saturation: $K_0=230^{+9}_{-11}$ MeV and $L=63^{+10}_{-13}$ MeV at 68$\%$ CL. They were obtained by making certain ad-hoc assumptions regarding their correlation with higher order coefficients, the skewness $J_0$ and the curvature $K_{sym}$ parameters respectively, see ~\secref{compmethod}. The latter was a posteriori justified given that the SE term mainly probes densities close to saturation, see \figref{denssens_lsymprot}, and its value is fixed below saturation $S$(0.1 fm$^{-3}$)=25.5 MeV. For SNM matter the sensitivity in the vicinity of the saturation point vanishes but becomes stronger close to $2 \rho_0$ with increasing impact energy, see ~\figref{denssens_k0prot}. Consequently, the quantity probed is a linear combination of $K_0$ and $J_0$ (depending on the average density probed), which may be in conflict with the assumed correlation between them. 

Existing independent constraints for SNM below $\rho_0$ can be used to disprove such a possibility. Certain static properties of nuclei, such as radius and mass, probe on average densities close to $\rho_c$=0.11 fm$^{-3}$, called crossing-density following the empirical observation that predictions for the EoS using available functionals are (nearly) the same in that region. It was subsequently shown that giant monopole resonances in an isotopic chain probe the third derivative of the energy density at the crossing-density, labeled $M$, rather than the compressibility modulus at saturation~\cite{Khan:2012ps,Khan:2010mv}. It can be conveniently determined from the expression $M=3\rho K'(\rho) |_{\rho=\rho_c}$, where $K'$ is the derivative, with respect to density, of compressibility $K(\rho)$. Comparison with experimental data for the centroid of giant monopole resonances (GMR) in $^{120}$Sn and $^{208}$Pb leads to $M$=1100$\pm$70 MeV.

The result of this study for SNM implies $M=1110^{+40}_{-50}$ MeV at 68$\%$CL, in excellent agreement with the GMR value. It would be of interest to redo the analysis reported in \secref{fitv1v2} with $K_0$ and $J_0$ varied independently, particularly since the GMR result implies a fine-tuning of $K_0$ rather than $J_0$. Preliminary work in this direction, to be reported elsewhere, indicates a correlation between the two parameters that indeed favors a moderately stronger dependence of $J_0$ on $K_0$ than assumed in this work (see ~\secref{compmethod}), but with minor impact on the density dependence of EoS of SNM in the probed range.

The density dependencies of EoS of SNM and SE up to twice saturation density determined in this study are presented in the left and right panels of \figref{eos_snm_se} respectively. Comparisons to previous studies based on HIC data, rapidity dependent elliptic flow of free protons and light clusters for SNM~\cite{LeFevre:2015paj} and neutron-to-proton or neutron-to-charged clusters elliptic flow ratios for SE~\cite{Russotto:2016ucm,Cozma:2017bre}, are shown. The present result for SNM is compatible with findings using IQMD to describe the $v_{2n}$ observable for proton, deuteron, triton and $^3$He, but much more accurate. Including rapidity dependent data outside the central rapidity region, as is the case with $v_{2n}$, softness the EoS, which may explain the difference in the central values of the two studies. Additionally, to our best knowledge, vacuum elastic $NN$ cross-sections are used in IQMD which is another potential source of differences (see also the discussion about the Cugnon parametrization of elastic cross-sections at the end of \secref{inemedelcs}).

The SE result of this study is compared with analyses of two dedicated experimental measurements accomplished by the FOPI-LAND~\cite{FOPI:1993wdf,Russotto:2011hq,Cozma:2013sja} and ASYEOS~\cite{Russotto:2016ucm} collaborations. Ratios of elliptic flows of isospin partners are better suited for the study of SE as sensitivity to isoscalar model parameters and systematic uncertainties of experimental data related to the reaction plane determination are significantly reduced. This explains, in part, the better accuracy of the ASYEOS result~\cite{Russotto:2016ucm} compared to this study. Adding the less accurate neutron-to-proton elliptic flow ratios determined by the FOPI-LAND collaboration has allowed the extraction of both $L$ and $K_{sym}$ parameters from experimental data, at the expense of larger uncertainties above 1.5$\rho_0$ using a previous version of dcQMD~\cite{Cozma:2017bre}. The accuracy of the present study is in-between the two extremes: it about twice less accurate at 68$\%$ CL compared to the ASYEOS analysis but the 95$\%$ CL constraint is more accurate than the combined FOPI-LAND and ASYEOS constraint. 

Results of a restricted set of ab-initio calculations employing different techniques but all relying on chiral interactions at various orders in the effective expansion, as indicated by label, with or without $\Delta$(1232) degrees of freedom are also displayed in \figref{eos_snm_se}: coupled-clusters (CC N$^2$LO$_{sat}$)~\cite{Ekstrom:2015rta}, many-body perturbation theory (MBPT N$^3$LO)~\cite{Drischler:2020yad}, self-consistent Green's functions approach (SCGF N$^3$LO)~\cite{Carbone:2019hym} and Brueckner-Hartree-Fock (BHF N$^3$LO$\Delta$/N$^2$LO$\Delta$)~\cite{Logoteta:2016nzc}. Predictions for the density dependence of SE energy are shown only for the last two models. The N$^2$LO$_{sat}$ interacting is fixed by fitting radii and binding energies of selected nuclei up of mass A=25, which allows a reasonable description of saturation properties of SNM already at third order in the effective expansion~\cite{Hagen:2015yea}. Including $\Delta$ baryon degrees of freedom in the two-body and three-body interactions at fourth and respectively third order in perturbation theory facilitates a more precise description of the empirical location of the saturation point. It may appear that all ab-initio predictions are 
outside the 95$\%$ CL interval of HIC results for SNM. In fact, a careful evaluation of uncertainties of ab-initio calculations demonstrates that for the MBPT N$^3$LO model the empirical saturation point of SNM is slightly outside the 95$\%$ CL interval, while the 1$\sigma$ uncertainty in binding energy reaches approximately 5 MeV at twice saturation density~\cite{Drischler:2020yad}. For SE the situation is better, the 1$\sigma$ uncertainty at 2$\rho_0$ is close to 2.5 MeV. Finally, it was shown in the same study that all determined quantities, including $K_0$ and $L$, carry additional systematic uncertainties induced by the employed regularization scheme, in particular the cut-off parameter value, that are significantly larger by at least a factor of two than those quoted above. 

\begin{figure}
\includegraphics[width=0.495\textwidth]{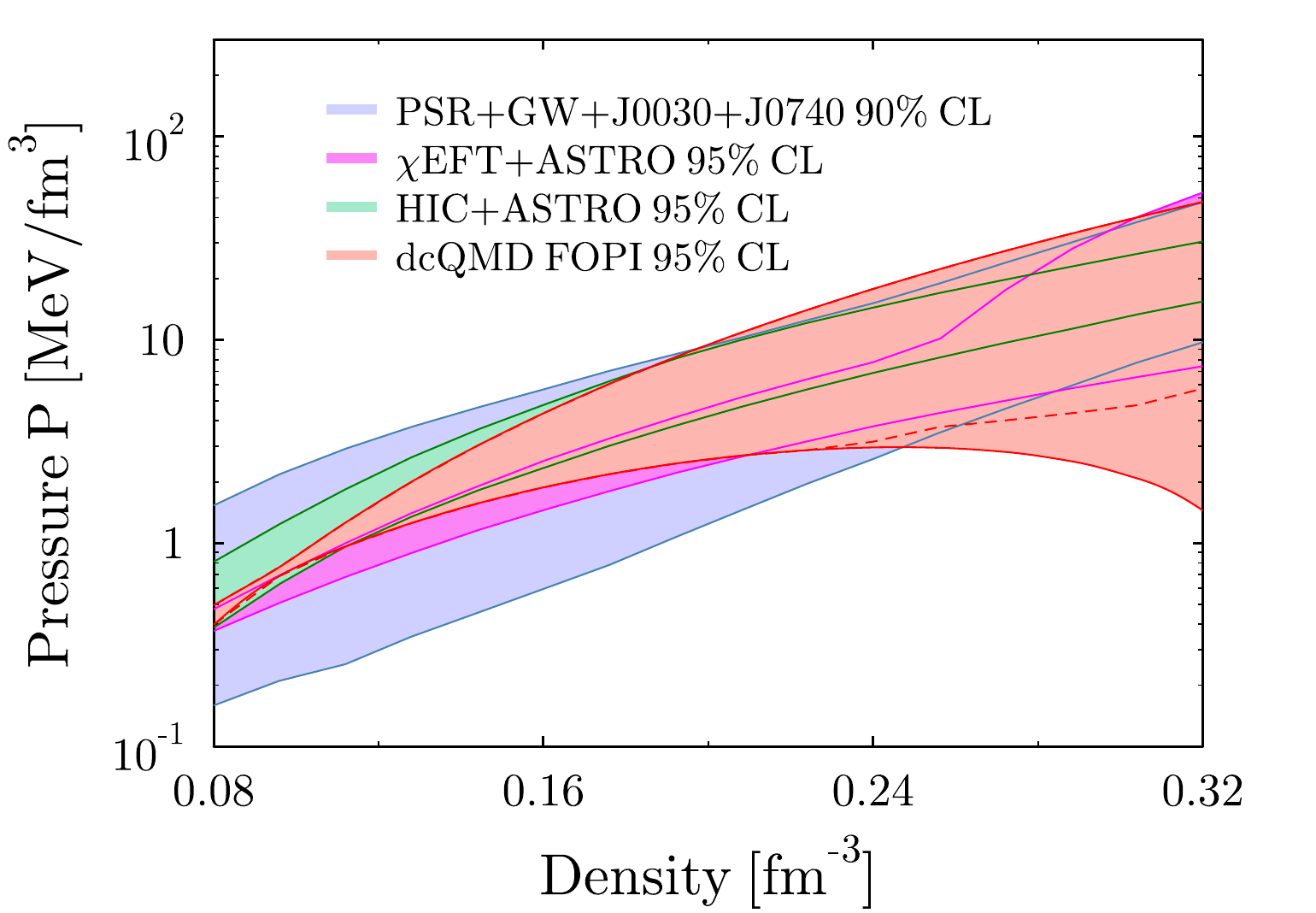}
\caption{\figlab{pressure_nsm} (Color online) Pressure as a function of density for cold neutron star matter ($\delta=0.93$) obtained in this study, compared with results from other three sources: astrophysics~\cite{Legred:2021hdx}, chiral perturbation theory~\cite{Huth:2021bsp} and nuclear physics measurements (structure and HICs)~\cite{Tsang:2023vhh}. The restriction of this study's result by imposing thermodynamics stability conditions is depicted by dashed curves, clearly visible only for the lower limit. }
\end{figure}

Astrophysical constraints are often presented by plotting pressure of cold neutron star matter (NSM) as a function of density. In \figref{pressure_nsm} the 95$\%$ CL result for pressure of nearly pure neutron matter ($\delta$=0.93) is shown using the constraints extracted in \secref{fitv1v2} for the correlation distribution of $K_0$ versus $L$. Results from three other sources, astrophysics ~\cite{Legred:2021hdx}, a combination of nuclear structure and HIC derived constraint~\cite{Tsang:2023vhh} and ab-initio quantum Monte-Carlo calculations based on two and three-body chiral nuclear forces~\cite{Huth:2021bsp} are shown for comparison. The astrophysics result relies on experimental detection of gravitational waves emitted in the process of neutron-star mergers and indirect measurements of radii and maximum masses of neutron stars, being the most uncertain of the four at densities above 1.5$\rho_0$. Both nuclear physics and ab-initio calculations constraints rely on a minimum sets of astrophysics input (maximum allowed masses for NS), causality and thermodynamic stability of NS matter to reject EoS's with either very soft or very stiff behavior at high densities. For comparison, the impact of thermodynamic stability ($\partial P /\partial\rho \ge 0$) for the dcQMD FOPI constraint is shown by dashed curves. The impact is visible only on the lower limit at densities above 1.5$\rho_0$. The dcQMD result is most accurate in the region slightly above half-saturation density as a consequence of fixing the symmetry energy at $\rho$=0.10 fm$^{-3}$ and of the imposed, but a posteriori proven realistic, correlation between $K_0$ and $J_0$. The impact of varying S($\rho$=0.10 fm$^{-3}$) on pressure should be assessed in a future study. The NSM EoS obtained in this study agrees well with the other three constraints over the entire range of densities below 2$\rho_0$ and particularly so below saturation where it is most accurate. 

Adding FOPI-LAND, ASYEOS and the very recent S$\pi$RIT~\cite{S:2023gzo} data sets for nucleonic observables to the list of fitted observables has the potential to further increase accuracy of SE and will be pursued in a future study. Pionic observables represent a promising addition for improving accuracy of constraints in the vicinity and above 1.5$\rho_0$. To make good use of the full potential of existing \cite{SpiRIT:2021gtq,HADES:2020ver} and near future available experimental data sets for this type of observables, planned to be measured by the S$\pi$RIT and HADES collaborations, dcQMD has to be improved in several ways: channels allowing non-resonant pion production and nucleonic resonance decay into final states containing two pions have to be included either from scratch or in a manner consistent with momentum dependent interactions. They are crucial for a realistic description of pion emission close to threshold and above 1 GeV/nucleon impact energy respectively over the entire range of measured rapidities and transverse momenta of multiplicity spectra.

\subsection{In-medium cross-sections}
\seclab{inmedcs}

\begin{figure*}[htb]
\includegraphics[width=0.495\textwidth]{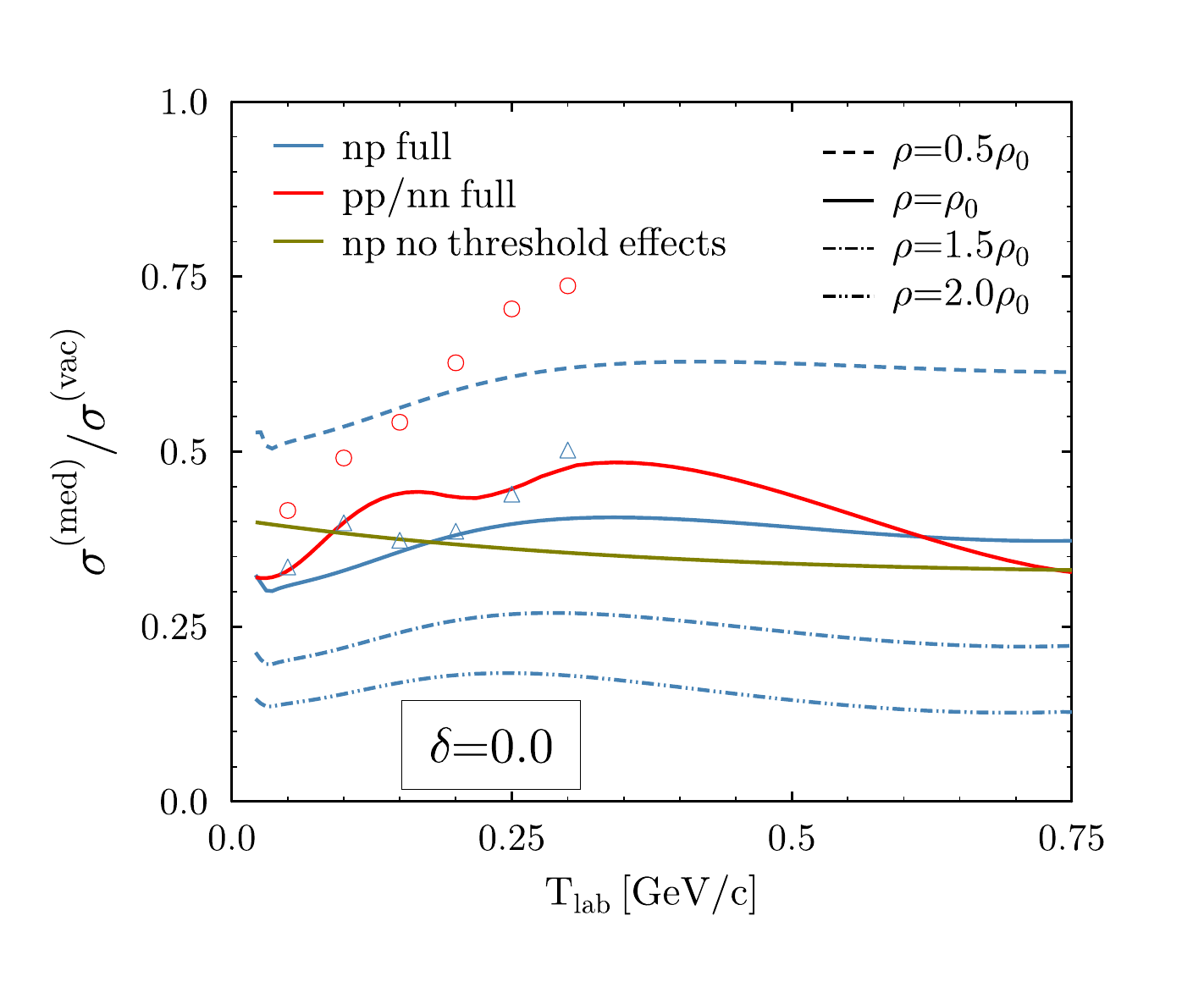}
\includegraphics[width=0.495\textwidth]{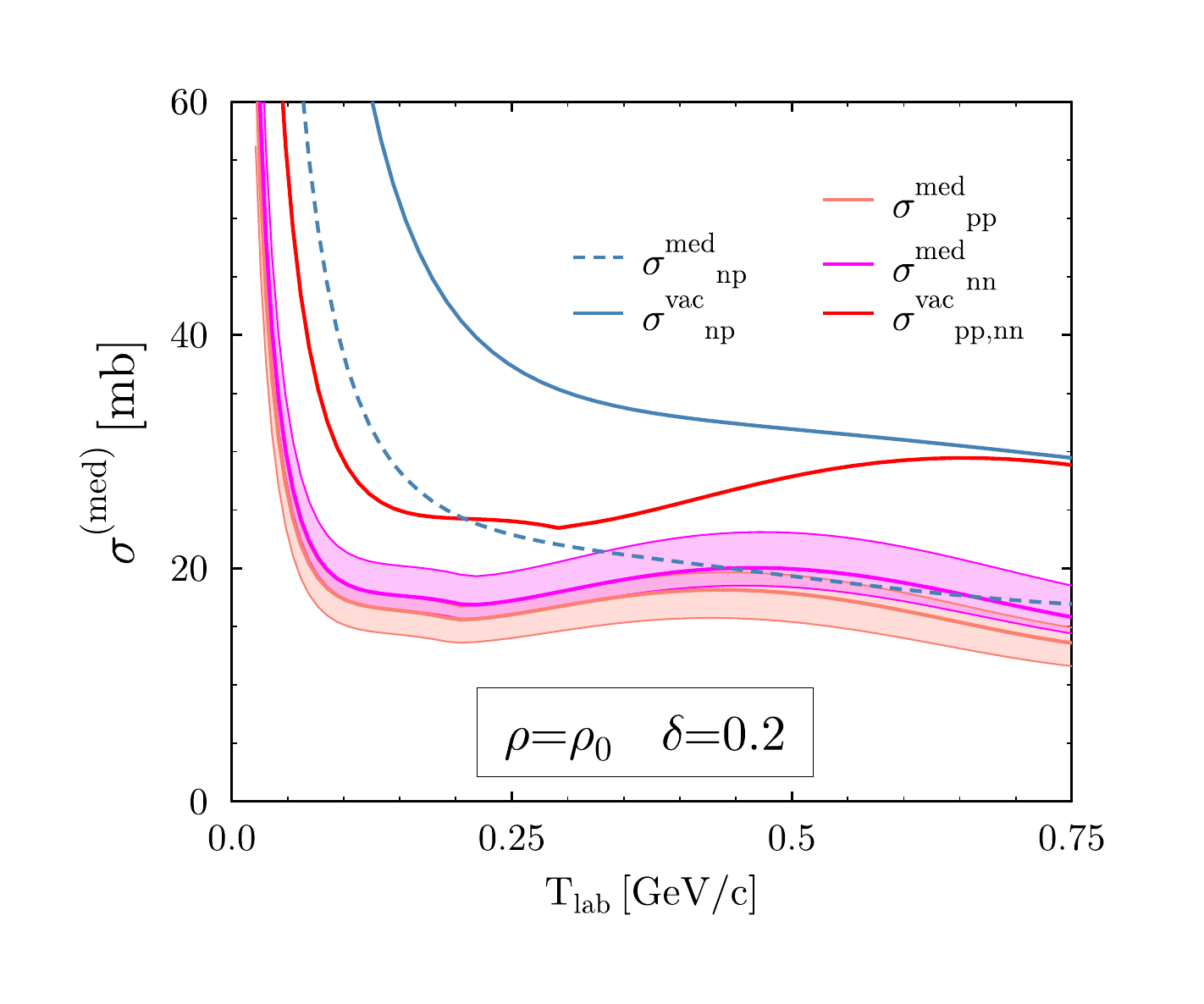}
\caption{\figlab{medcs} (Color online) (Left Panel) In-medium modification factor of elastic $NN$ cross-sections in cold (T=0.0 MeV) symmetric nuclear matter of specified density as function of the kinetic energy of the pair $T_{lab}$. Open triangles and open squares represent result of microscopical calculations at saturation density for $np$~\cite{Li:1993ef} and $pp$~\cite{Li:1993rwa} elastic scattering respectively. (Right Panel) In medium elastic $NN$ cross-sections in cold ANM ($\delta$=0.2) at saturation density as function of $T_{lab}$. Bands indicate the 68$\%$ CL due to uncertainties of $\Delta m^*_{np}$ and $\beta_2$ parameters; $\sigma^{med}_{np}$ does not dependent on $\beta_2$ and the sensitivity to $\Delta m^*_{np}$ is negligibly small being of second order. In all cases the total momentum of the two-nucleon pair in the rest frame of NM is set to zero.}
\end{figure*}

In-medium modification of elastic $NN$ cross-sections is presented in \figref{medcs}. The left panel shows the ratio $\sigma^{(med)}_{np}$/$\sigma^{(vac)}_{np}$ in SNM at T=0 MeV as function of the kinetic energy of the colliding pair at zero total momentum for several densities. The suppression is stronger at higher densities and displays a mild dependence on the kinetic energy of the pair at the lower end of this variable.
The latter is seen to be induced by threshold effects as a result of the interplay between the energy dependence of transition amplitudes and momentum dependence of the interaction that induces a momentum dependence of effective masses. For $\rho=\rho_0$ the ratio $\sigma^{(med)}_{pp/nn}$/$\sigma^{(vac)}_{pp/nn}$ is also shown. Differences with respect to the $np$ case originate from the dissimilar energy dependence of vacuum transition amplitudes.

Microscopic calculations of in-medium elastic $NN$ cross-sections in SNM up to pion production threshold using the Dirac-Brueckner Hartree Fock approach have been reported in Ref.~\cite{Li:1993rwa,Li:1993ef} and are shown in the left panel of \figref{medcs} for saturation density. The in-medium reduction factor is similar to the one extracted from HIC observables for values of $T_{lab}$ close to 50 MeV for all isospin channels but displays a stronger energy dependence particularly in the $nn/pp$ channel overshooting the values obtained in this work at $T_{lab}$=300 MeV. A possible explanation of this behavior is related to the minimalistic in-medium modification of transition amplitudes adopted in this work: they were assumed to depend only on density and isospin asymmetry, a potential momentum dependence induced by Pauli blocking of intermediate states has not been accounted for. A similar calculation performed by a different group~\cite{Fuchs:2001fp} exhibits stronger suppression of cross-sections at low kinetic energies and a smaller reduction relative to vacuum values above Fermi momentum. This is due to a modification of the optical theorem induced by Pauli blocking of intermediate states which also requires a revision of the expression for cross-sections. Such effect has not been account for in Refs.~\cite{,Li:1993rwa,Li:1993ef} or the present study. 

\begin{figure*}[htb]
\includegraphics[width=0.495\textwidth]{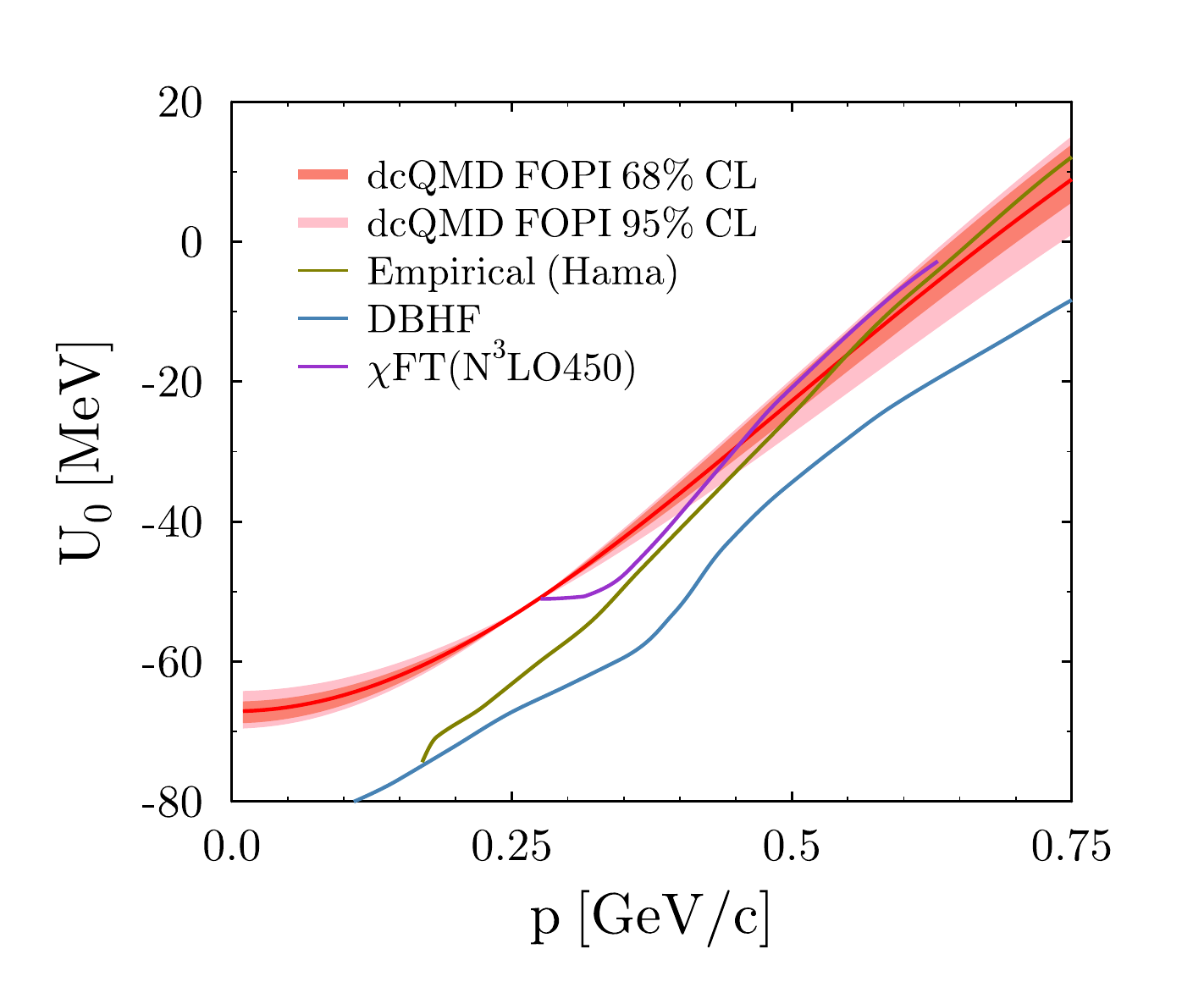}
\includegraphics[width=0.495\textwidth]{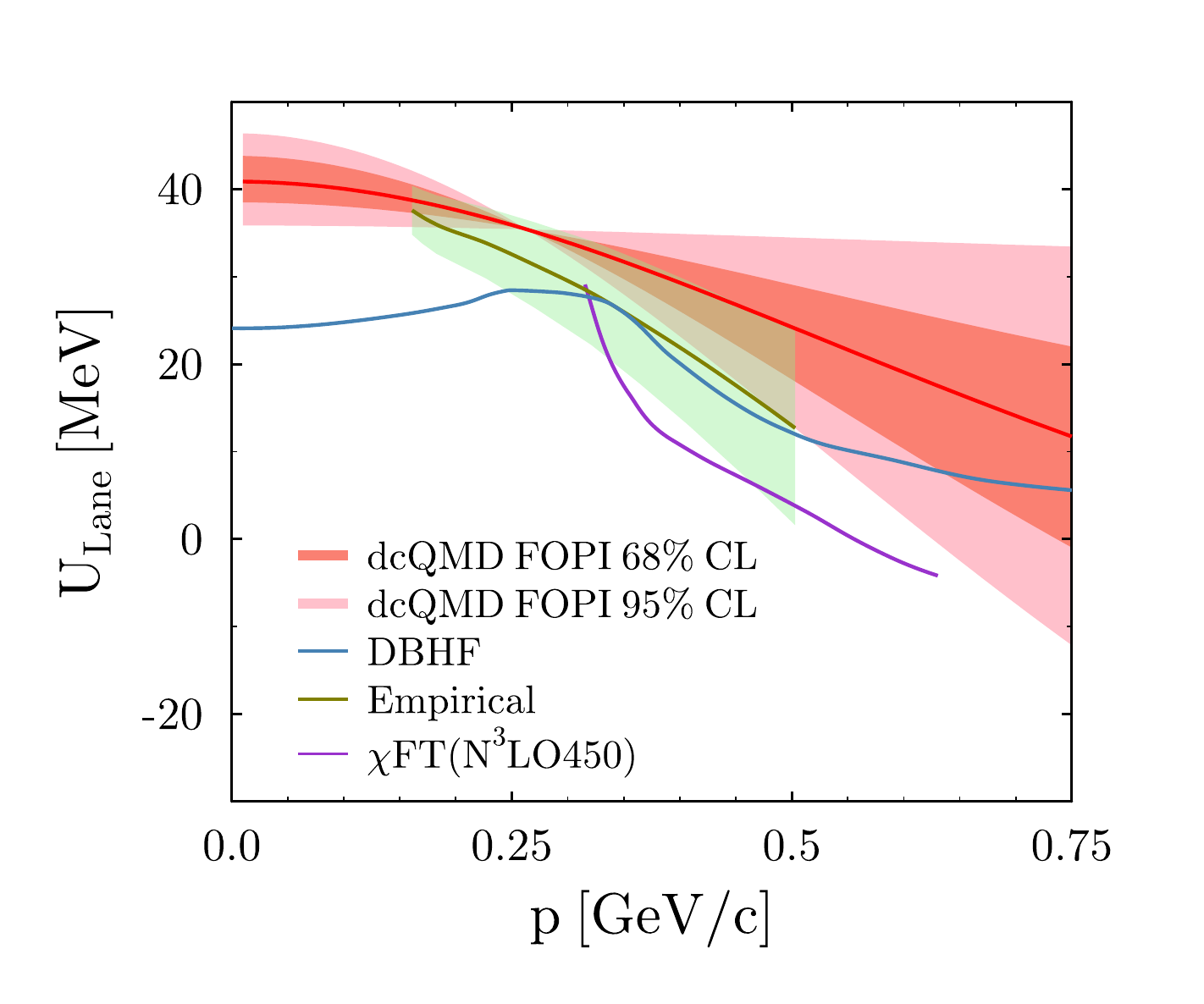}
\caption{\figlab{optpot} (Color online)(Left Panel) Isoscalar optical potential at saturation density as a function of nucleon momentum from four sources: HIC (this study), empirical Hama potential~\cite{Hama:1990vr}, microscopical DBHF ~\cite{vanDalen:2005sk} and $\chi$FT~\cite{Holt:2015dfa} calculations (Right Panel) The same as in the left panel but for the symmetry (Lane) potential. The shown empirical Lane potential depicts a parametrization~\cite{Li:2004zi} of analyses of nucleon-nucleus scattering experiments at beam energies below 100 MeV.}
\end{figure*}

At densities above saturation, microscopic calculations favor a stronger reduction only for low kinetic energies of the colliding pair, more so for $np$ than $nn/pp$. Outside that region the reduction factor shows either a small dependence ($np$) or an increase ($nn/pp$) with density towards vacuum values which are reached at twice saturation density for the latter case. Clearly, the simple parametrization for in-medium modification of transition amplitudes together with the monotonic dependence of effective masses on density cannot mimic such a complex behavior. A sensitivity study, similar to the one described in ~\secref{probeddens} for the EoS, shows that flow observables are sensitive to in-medium cross-sections in a range between half and twice saturation density with a maximum in the vicinity of 1.25$\rho_0$. This offers support for the observed better agreement between the HIC and microscopical model results at densities close to saturation.

The right panel of \figref{medcs} presents in-medium elastic $NN$ cross-sections at saturation density for isospin asymmetric nuclear matter ($\delta$=0.2) compared to vacuum ones. A stronger suppression at lower kinetic energies is observed, which also exhibits a stronger energy dependence compared to isospin symmetric matter. A small splitting of $nn$ and $pp$ cross-sections, $\sigma^*_{nn}>\sigma^*_{pp}$, is favored by HIC data, which is however compatible with no splitting already at 68$\%$ CL. Microscopic calculations generally favor a positive splitting of $pp$ and $nn$ cross-sections at saturation in isospin asymmetric matter~\cite{Sammarruca:2005tk,Dong:2010fxi,Sammarruca:2013bda}. A variation of its sign with density has been reported~\cite{Sammarruca:2005tk}, with $\sigma^*_{nn}>\sigma^*_{pp}$ at densities close to $\rho_0/2$ for certain values of the relative momentum of the colliding pair. A positive  splitting of $pp$ and $nn$ cross-sections in dcQMD can be enforced by either choosing negative values for $\Delta m^*_{np}$ or positive ones for $\beta_2$  enough to compensate the tendency in the opposite direction of other parameters, mostly $\beta_2$ and $\Delta m^*_{np}$ respectively.

\subsection{Momentum dependent potentials}
\seclab{mdi}

A comparison of the momentum dependence of the optical potential at saturation density deduced in this work with results of microscopic calculations~\cite{vanDalen:2005sk,Holt:2015dfa} and empirical information~\cite{Hama:1990vr,Cooper:1993nx,Li:2004zi,Xu:2014cwa} is presented in \figref{optpot}. The left and right panels show the isoscalar and the isovector (Lane) components respectively. A good agreement between the HIC isoscalar optical potential and the empirical (Hama) one is observed for larger values of momenta. A moderate discrepancy exists for momenta close to Fermi momentum and below. A better agreement can be obtained by setting the effective isoscalar mass to $m^*$=0.55 m$_N$, in strong disagreement with the value obtained for this parameter in~\secref{fitv1v2}. In contrast, DBHF calculations are closer to the empirical Hama potential at low momenta and start to deviate above Fermi momentum, being generally more attractive than the HIC result. The $\chi$FT isoscalar optical potential~\cite{Holt:2015dfa} agrees well with both HIC and empirical result over the restricted range of momenta for which calculation are available.

A good agreement between the HIC and the empirical results for the momentum dependence of the Lane potentials is observed, see right panel of \figref{optpot}. The decrease of its repulsion with increasing momentum is due to a positive value for $\Delta m^*_{np}$. A negative value for this parameters, as suggested by several analyses of HIC observables using the ImQMD transport model~\cite{Morfouace:2019jky,SpRIT:2023als}, would lead to an opposite behavior, in stark contrast with the empirical Lane potential. It was shown in \secref{fitv1v2} that the inclusion of threshold effects in elastic nucleon-nucleon collisions is crucial for obtaining a positive neutron-proton effective mass splitting. Microscopic DBHF calculations~\cite{vanDalen:2005sk} are in good agreement with both HIC and empirical results but deviate from the former at very low momenta. The $\chi$FT Lane potential~\cite{Holt:2015dfa} is generally less repulsive but still in good agreement with the empirical one.

\section{Summary and Conclusions}
\seclab{sumcon}
The dcQMD transport model is employed to study several aspects of the in-medium nucleon-nucleon interaction, namely in-medium modification of elastic cross-sections, momentum dependence of the optical potential and equation of state of nuclear matter. This study is a first step towards making use of the extensive database of heavy-ion collisions results to its full potential for the purpose of extracting accurate information on these quantities. To that end, theoretical predictions are compared to experimental data for stopping observable $varxz$, directed and elliptic flows at mid-rapidity measured by the FOPI collaboration for intermediate energy HIC. In order to keep the analysis close to the simplest possible, the upper limit of the beam energy range is restricted to 800 MeV/nucleon with two unimportant exceptions. Below this limit the fraction of nucleons excited into resonances, mostly $\Delta(1232)$, in mid-central collisions is less than 5$\%$. Consequently, the impact of the poorly known resonance potentials on nucleonic observables such as collective flows is minute. As the impact energy is raised above 1.0 GeV/nucleon resonance potentials affect nucleonic observables by a non-negligible amount, requiring an extension of the analysis to include pionic observables. Owing to an outstanding discrepancy between FOPI and HADES collaborations results for pion multiplicities and the present omission in dcQMD of resonance decay channels with two pions in the final state, inclusion of experimental data at impact energies of 1.0 GeV/nucleon and above is postponed for a future study.

Several ingredients of the model are improved. Firstly, accurate parametrizations of vacuum elastic nucleon-nucleon cross-section experimental data are built and used. Previously used parametrizations, such as the well known Cugnon one, include an effective density and isospin asymmetry independent in-medium modification at low impact energies, particularly in the neutron-proton channel. Instead, we use an ad-hoc change of transition amplitudes by a multiplicative factor that depends on density, isospin asymmetry and isospin projections of the colliding particles, supplemented by threshold effects. In-medium modification of differential cross-section guided by results of microscopical calculations are also implemented. The surface term of the Pauli blocking algorithm is modified to account for the fact that high density is correlated with high temperature in heavy-ion reactions and consequently its strength diminishes above saturation. Finally, the coalescence algorithm is applied continuously during the time evolution of the reaction, leading to an early identification of clusters at the local freeze-out time, rather than at the end of the reaction as in previous versions of the model. This approach corrects spurious emission of free nucleons in QMD type models and leads to a visibly improved description of light cluster (H ans He isotopes) multiplicities in central collisions, in particular at impact energies of 400 MeV/nucleon and above.

Varied parameters of the model are determined by a fit to the following set of experimental data measured by the FOPI collaboration: proton, deuteron and triton $varxz$ for 14 systems of various isospin content with impact energies between 150 and 1000 MeV/nucleon; rapidity directed flow for $Z=1$ and $Z=2$ fragments at 250 and 400 MeV/nucleon beam energy in mid-central NiNi, XeCsI and AuAu collisions; rapidity dependent directed ($p$, $d$, A=3 and $\alpha$ clusters) and elliptic flow ($p$, $d$ and $\alpha$ clusters) and transverse momentum elliptic flow ($p$, $d$ and $t$) in mid-central AuAu collisions of impact energy between 150 and 800 MeV/nucleon. Transverse momentum dependent directed flow spectra, while available experimentally, are not included in the fit since theoretical values for protons are contaminated by heavier clusters. This also represents the motivation to restrict rapidity dependent observables to the mid-rapidity part of spectra.

Eight free model parameters are determined employing a multivariate analysis that takes into account systematic uncertainties induced on model predictions by the coalescence afterburner. Three of them are used to adjust in-medium elastic nucleon-nucleon cross-sections in isospin symmetric and asymmetric matter. For the latter case, dependence on isospin asymmetry as well as a possible splitting of $nn$ and $pp$ cross-sections are considered. A suppression of elastic nucleon-nucleon cross-sections in symmetric nuclear matter at saturation density by about 60$\%$ relative to vacuum values is deduced, in qualitative agreement with microscopical calculations. A strong dependence of the suppression factor on isospin asymmetry is evidenced together with a positive splitting, though compatible with no effect already at 68$\%$ confidence level, of $nn$ versus $pp$ cross-sections in isospin asymmetric matter. Experimental data for light isospin symmetric systems (stopping in CaCa, NiNi, RuRu and to a lesser extent directed flow in NiNi) are found important for a precise determination of in-medium cross-sections and breaking of a degeneracy between parameters that control in-medium modification of cross-sections depending on density and isospin asymmetry that would occur if only neutron rich systems were considered.

Other three model parameters control the momentum dependence of the optical potential and are related to the isoscalar nucleon effective mass ($m^*$), neutron-proton effective mass splitting ($\Delta m^*_{np}$) and strength of the isoscalar potential at infinitely large momentum ($U_\infty$). The following values are deduced at 68$\%$ confidence level: $m^*=0.695^{+0.014}_{-0.018}$, $\Delta m^*_{np}=(0.17^{+0.10}_{-0.09})\delta$ and $U_\infty=104^{+12}_{-7}$  MeV. They lead to isoscalar and isovector (Lane) optical potentials that are in reasonable agreement with empirical constraints and microscopic calculations (both DBHF and $\chi$FT). Threshold effects are found to be crucial for obtaining a positive value for $\Delta m^*_{np}$ that agrees with the world average for this quantity. Fixing $U_\infty$ at less repulsive values used in the literature ($e.g.$ 75 MeV used in previous studies) has a significant impact on the extracted density dependence of the equation of state (EoS). The compressibility modulus of symmetric nuclear matter ($K_0$) and slope of the symmetry energy at saturation ($L$) are affected at the level of three and two standard deviations respectively. Using a momentum independent Lane potential ($\Delta m^*_{np}$=0.0) has very little impact on any of the other parameters with the exception of $m^*$ which is impacted at one standard deviation.

The remaining two parameters are used to vary compressibility modulus of symmetric nuclear matter ($K_0$) and slope of the symmetry energy at saturation ($L$). The symmetry energy is fixed at two thirds saturation density using a value extracted from nuclear structure experiments: $S(0.10~\rm{fm}^{-3})= 25.5$ MeV. The extracted constraint at 68$\%$ confidence level reads: $K_0=230^{+9}_{-11}$ MeV and $L=63^{+10}_{-13}$ MeV.
Inclusion of transverse momentum dependent elliptic flows are found crucial for obtaining a good accuracy. Applying the coalescence model at the local freeze-out time has a large impact on the favored value for $K_0$ but not $L$. In contrast, threshold effects are seen to have a sizable impact on $L$ but less so on $K_0$. The latter is related to a correlation between $K_0$ and in-medium modification of elastic cross-sections in symmetric nuclear matter, which is constrained by stopping observables that are less affected by threshold effects. Omitting $varxz$ from the fit results in strong impact of threshold effects on $K_0$ as well. The deduced density dependence of EoS of SNM and ANM compare well to existing empirical and first-principle theoretical results.

Lastly, the probed density is determined by computing the functional derivative of flow observables with respect to the equation of state. It is found that, qualitatively, all cluster species probe sub-saturation and supra-saturation densities equally well, however elliptic flow is better suited for the study of symmetry energy above saturation than directed flow. The same distinction does not hold true for the EoS of SNM. The sensitivity of flow observables to the EoS of SNM shows a maximum above saturation density whose location is correlated with beam energy. In particular, at 800 MeV/nucleon beam energy the maximum is located close to twice saturation density. This offers the prospect of studying this quantity at ever higher densities by increasing impact energy. In contrast, the sensitivity to symmetry energy shows no discernible dependence on beam energy. This difference in behavior is the result of the suppression of the drift term of the isovector force due to favored close to linear dependence of SE on density and sub-unitary value of isospin asymmetry. The dominant isovector diffusion component shows no such dependence as a result of dependence of the gradient of isospin asymmetry with beam energy that compensates the increase of symmetry energy with density. Consequently, studying the symmetry energy at higher densities will have to rely on finding more suitable observables and improving experimental accuracy rather than increasing beam energy.

\begin{acknowledgments}
The research of M.D.C. has been financially supported by the Romanian Ministry of Education and Research through Contract No. PN 23 21 01 01/2024 and NAFRO2 (part of the FAIR-RO programme 2024-2026). The authors aknowlegdes many useful discussions with members of the TMEP Collaboration. Computing resources have been partly provided by the Institute for Cyber-Enabled Research (ICER) at Michigan State University, USA.
\end{acknowledgments}

\setcounter{section}{1}
\setcounter{equation}{0}
\renewcommand\theequation{\Alph{section}.\arabic{equation}}
\section*{Appendix A}
\seclab{appendix}

Predictions for nucleonic observables and their comparison with experimental data are provided in this Appendix. Model parameters are set to the values extracted in \secref{fitv1v2} from a fit of $varxz$, directed and elliptic flows of nucleonic observables to experimental data, see table \tabref{fitflow8par}. Comparison to the experimental data sets FOPI1 and FOPI2, see \tabref{expdataset} for details of kinematic restrictions, are extended away from mid-rapidity up to $y/y_P \leq 1.0$. Predictions for transverse momentum dependent $v_1$, which are available, but were not used in \secref{fitv1v2} due to contamination of theoretical proton spectra by clusters (see \figref{fopi2_v1pt}), are also provided.

Comparison of theoretical values for $varxz$ for protons, deuterons and tritons to their experimental counterpart is provided in \figref{fopi_varxz}. Theoretical values display uncertainties comparable to the experimental ones as result of variation of coalescence parameters $\delta r$ and $\delta p$. Results for directed flows for $Z=1$ and $Z=2$ fragments are presented in \figref{fopi1_v1rp} and \figref{fopi1_v1pt}. None of the data points of the latter set, comprising of transverse momentum directed flow spectra was not used in the fit. Results for directed and elliptic flows of (depending on either rapidity or transverse momentum) of individual cluster species are show in \figref{fopi2_v1rp} (rapidity dependent $v_1$), \figref{fopi2_v1pt} (transverse momentum dependent $v_1$), \figref{fopi2_v2rp} (rapidity dependent $v_2$) and \figref{fopi2_v2pt} (transverse momentum dependent $v_2$). Also in this case, transverse momentum dependent $v_1$ spectra were not used to fit model parameters. Uncertainties of flow predictions are depicted as bands in the mentioned figures. They are given by the sum, in quadrature, of statistical and systematic uncertainties. The latter is the result of variation of coalescence parameters.

\begin{figure*} [htb]
\includegraphics[width=0.975\textwidth]{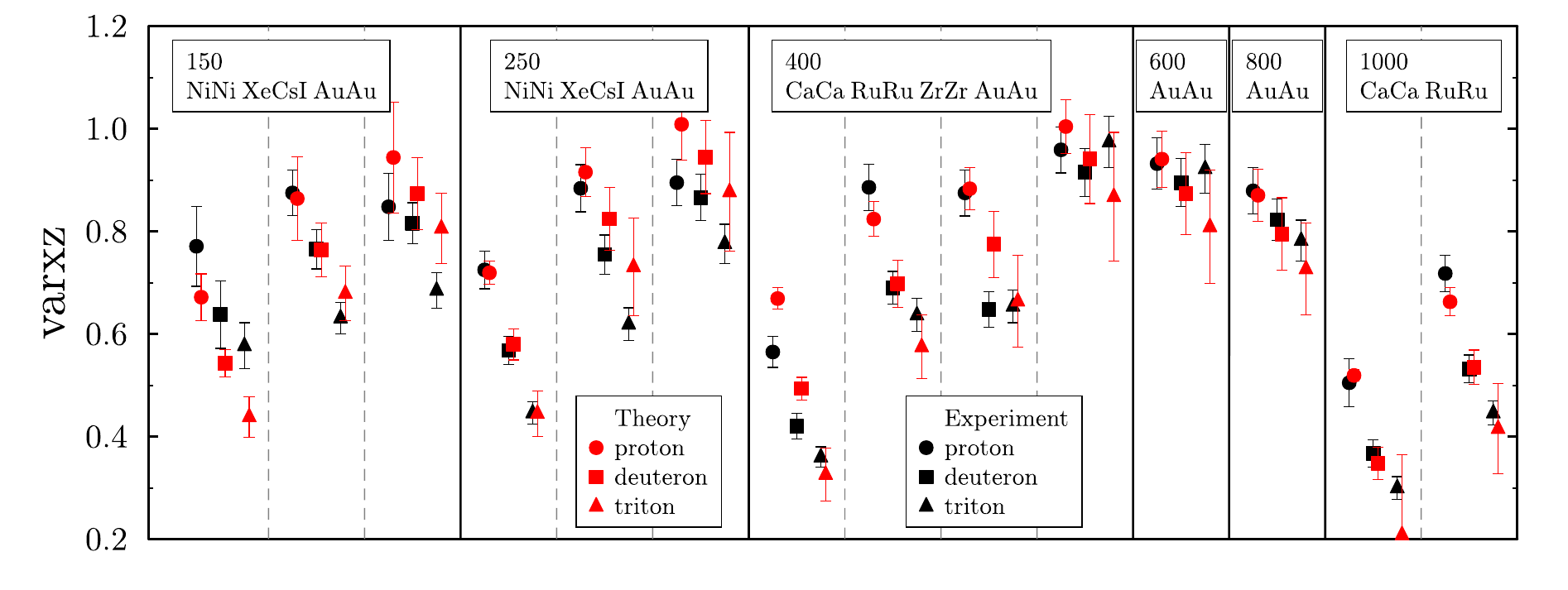}
\caption{\figlab{fopi_varxz} (Color online) Comparison of theoretical predictions for proton, deuteron and triton $varxz$ to experimental data~\cite{FOPI:2010xrt} for 14 selected systems.}
\end{figure*}

\begin{figure*} [htb]
\includegraphics[width=0.925\textwidth]{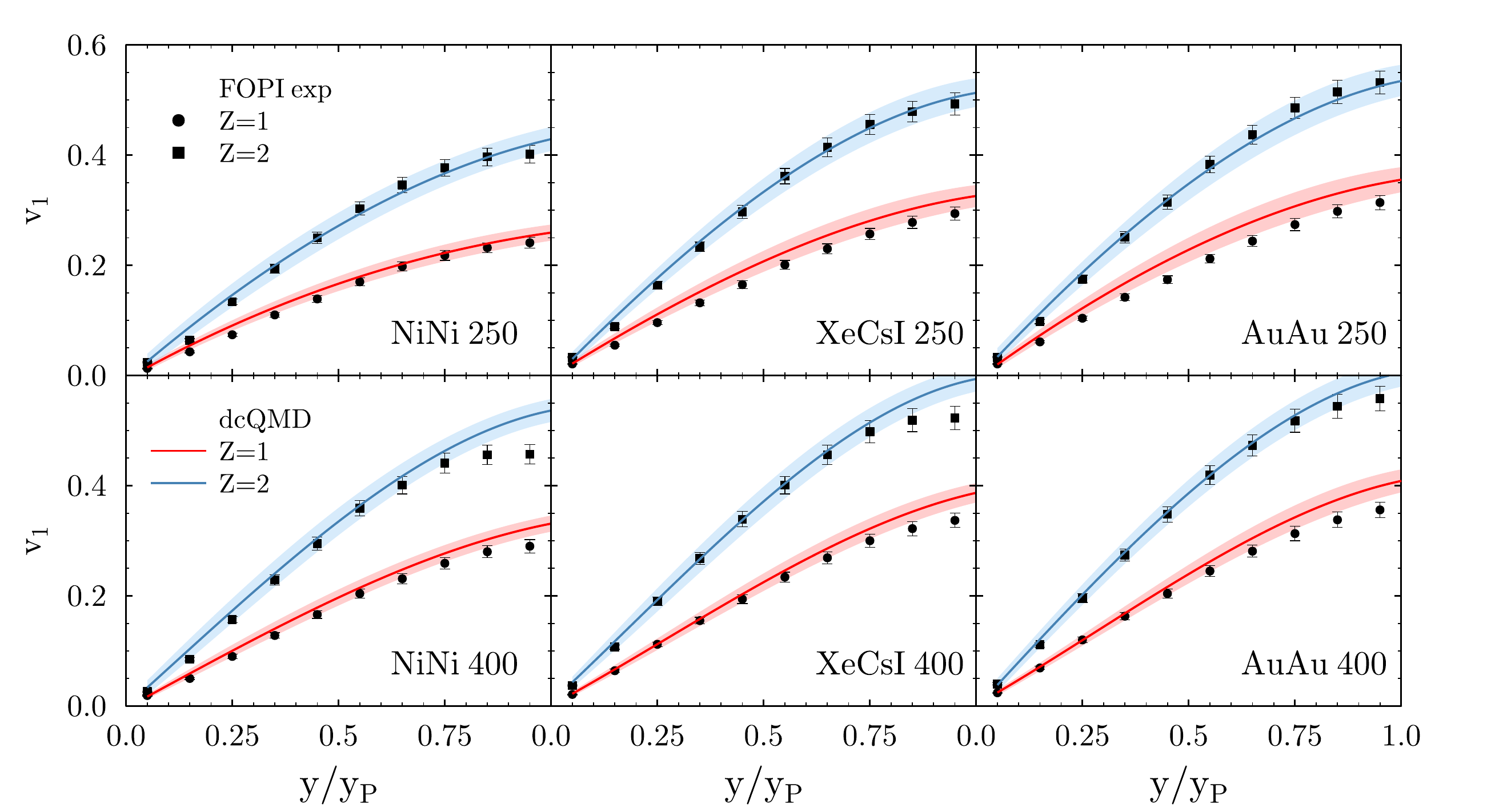}
\caption{\figlab{fopi1_v1rp} (Color online) Theoretical predictions for rapidity dependent $v_1$ of $Z=1$ and $Z=2$ fragments are compared to experimental data ~\cite{FOPI:2004bfz}, see ~\tabref{expdataset} for additional details. Only data points with $y/y_P\leq 0.5$ have been included in the fit described in ~\secref{fitv1v2}. Labels in each panel specify the system and impact energy in MeV. Bands signify theoretical uncertainty which is the sum, in quadrature, of statistical and systematic uncertainties. The latter is due to variation of coalescence algorithm parameters, see~\secref{fitv1v2}.}
\end{figure*}

\begin{figure*} [htb]
\includegraphics[width=0.925\textwidth]{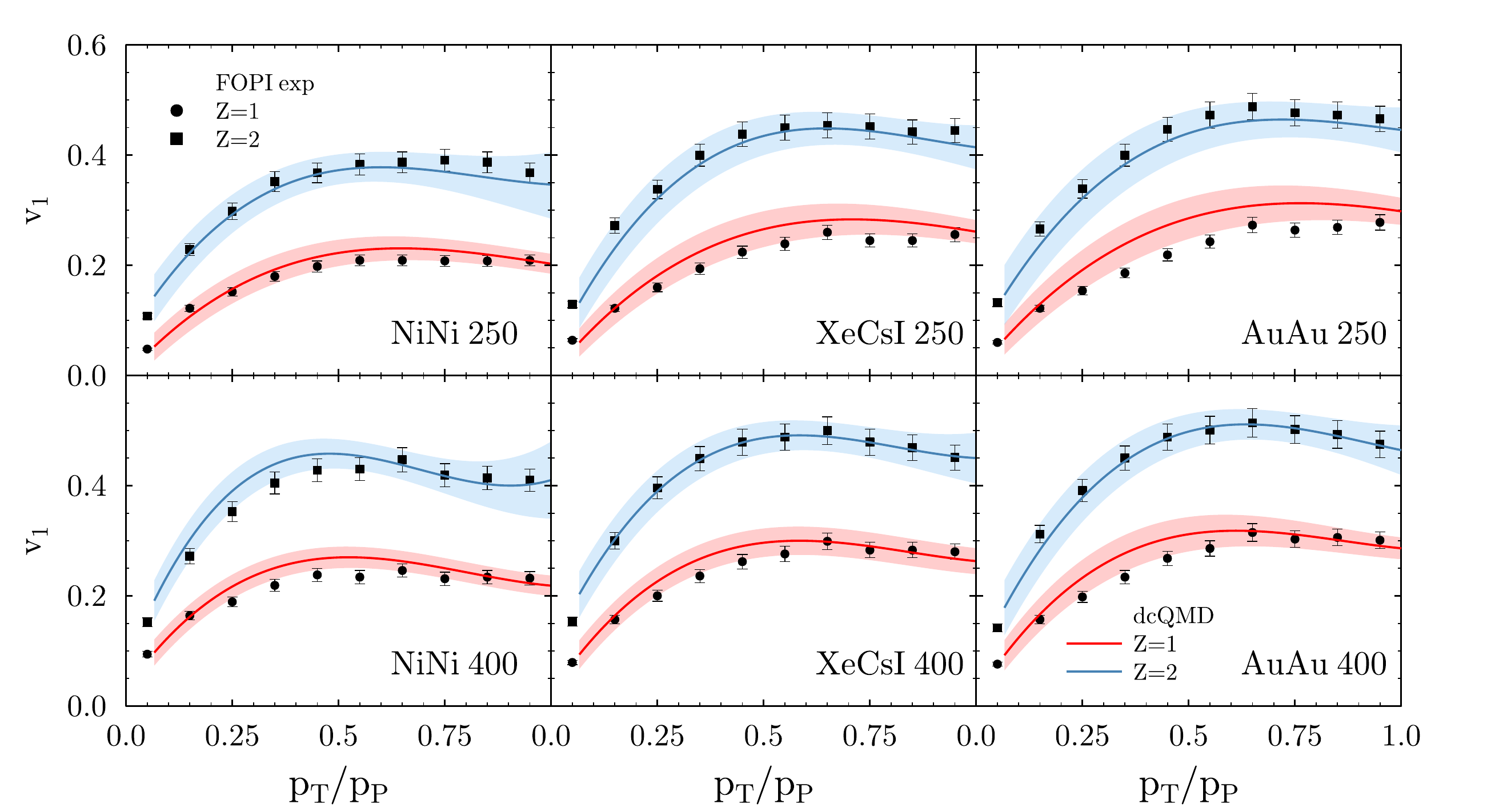}
\caption{\figlab{fopi1_v1pt} (Color online) Theoretical predictions for transverse momentum dependent $v_1$ of $Z=1$ and $Z=2$ fragments are compared to experimental data ~\cite{FOPI:2004bfz}, see ~\tabref{expdataset} for additional details. \underline{None} of the data points have been included in the fit described in ~\secref{fitv1v2}. The same comments as for \figref{fopi1_v1rp} apply.}
\end{figure*}

\begin{figure*} [htb]
\includegraphics[width=0.925\textwidth]{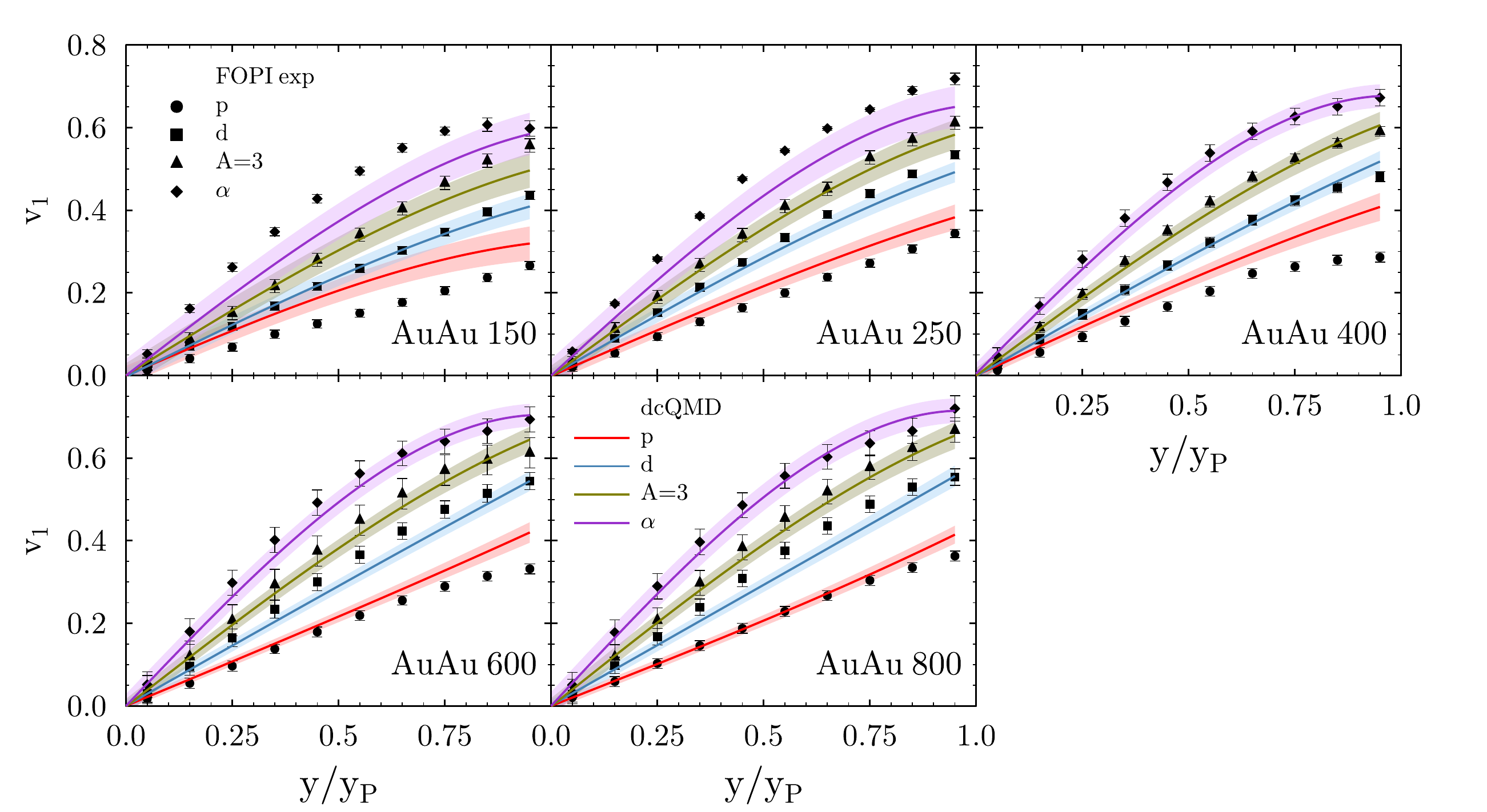}
\caption{\figlab{fopi2_v1rp} (Color online) Theoretical predictions for rapidity dependent $v_1$ of protons, deuterons, $A=3$ and $\alpha$ clusters are compared to experimental data ~\cite{FOPI:2011aa}, see ~\tabref{expdataset} for additional details. Only data points with $y/y_P\leq 0.5$ and $T_{lab}\geq 250$ MeV have been included in the fit described in ~\secref{fitv1v2}. The same comments as for \figref{fopi1_v1rp} apply.}
\end{figure*}

\begin{figure*} [htb]
\includegraphics[width=0.925\textwidth]{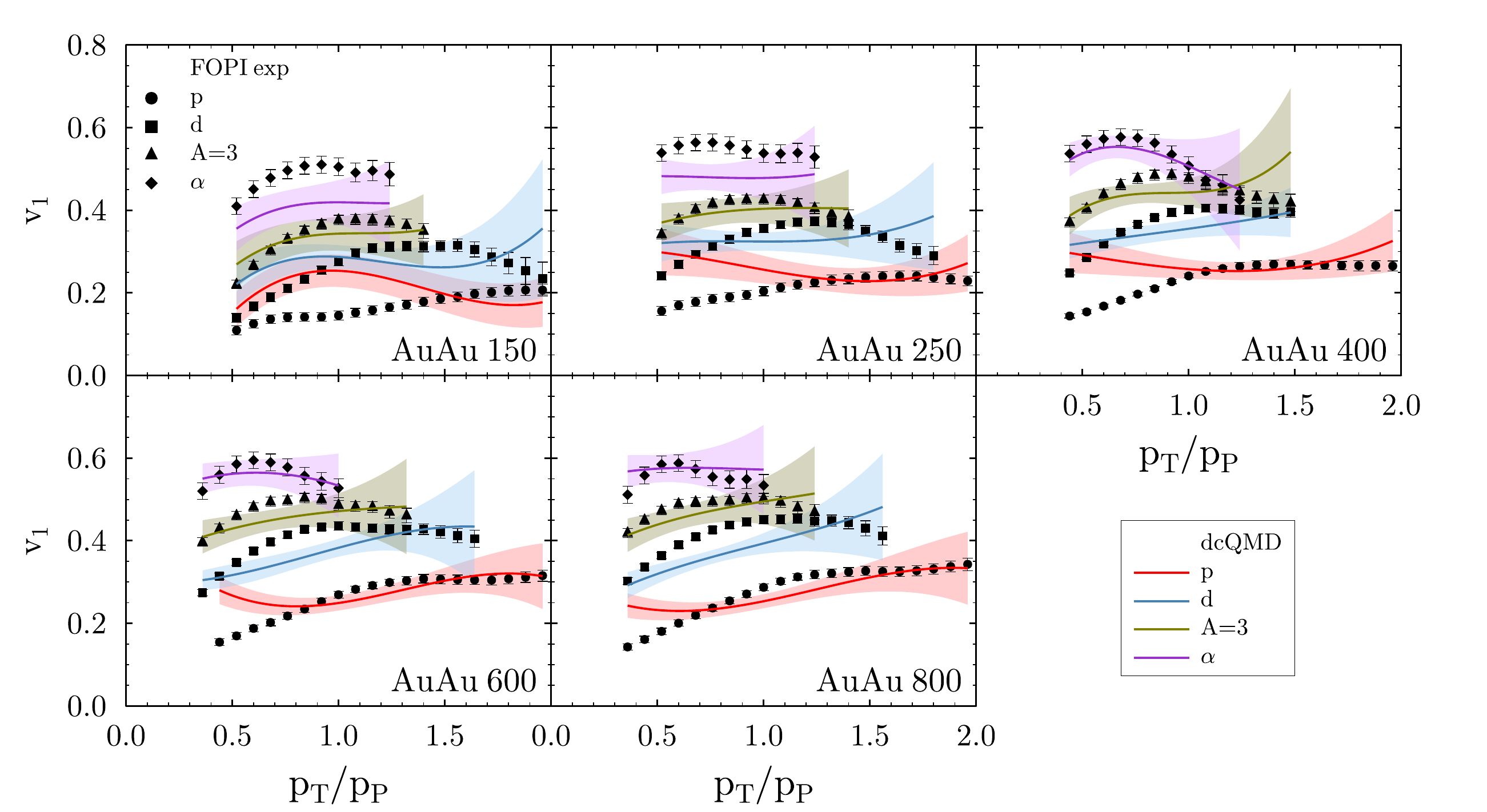}
\caption{\figlab{fopi2_v1pt} (Color online) Theoretical predictions for transverse momentum dependent $v_1$ of protons, deuterons, $A=3$ and $\alpha$ fragments are compared to experimental data ~\cite{FOPI:2011aa}, see ~\tabref{expdataset} for additional details. \underline{None} of the data points have been included in the fit described in ~\secref{fitv1v2}. The same comments as for \figref{fopi1_v1rp} apply.}
\end{figure*}

\begin{figure*} [htb]
\includegraphics[width=0.925\textwidth]{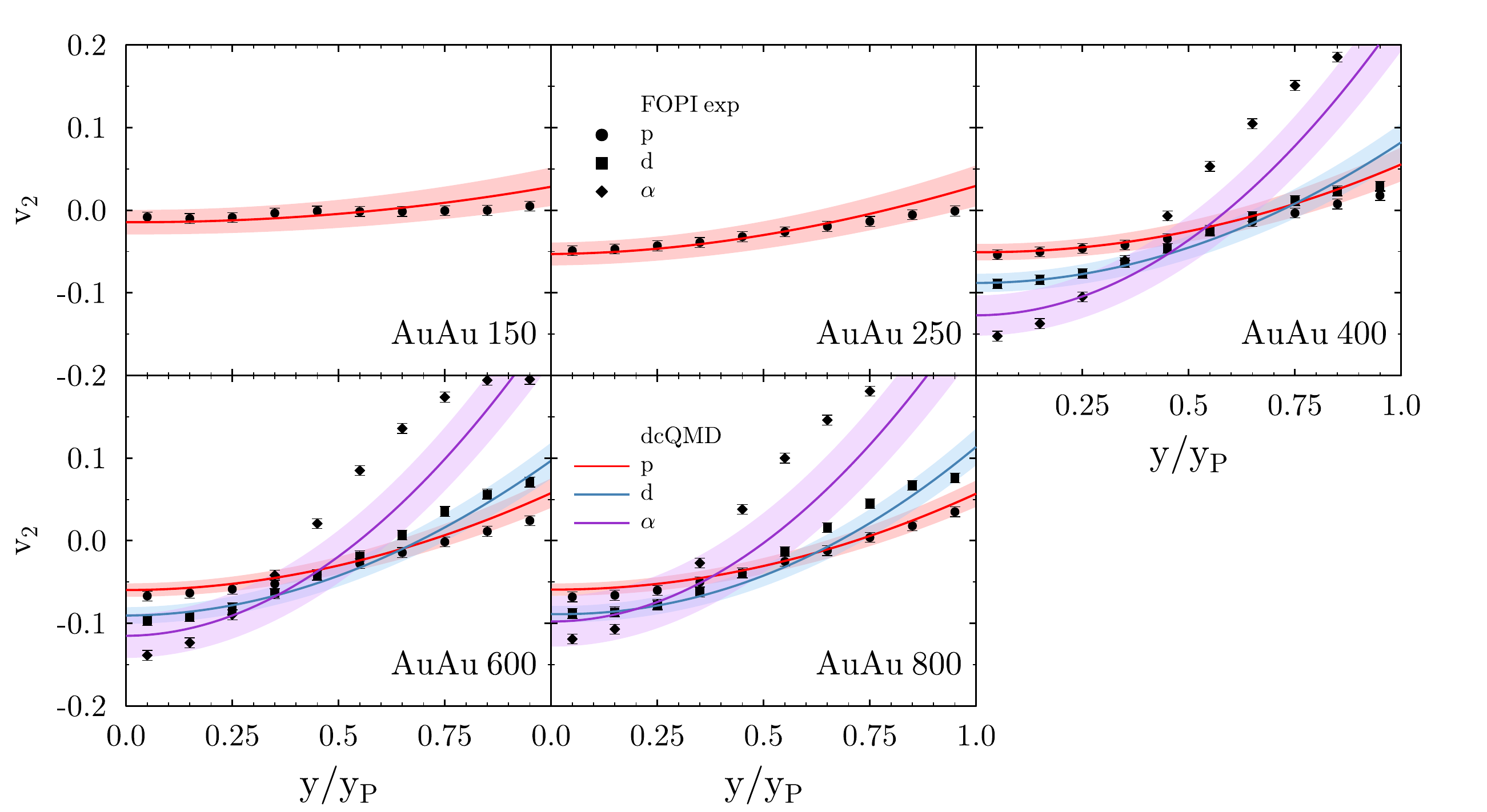}
\caption{\figlab{fopi2_v2rp} (Color online) Theoretical predictions for rapidity dependent $v_2$ of protons, deuterons and $\alpha$ clusters are compared to experimental data ~\cite{FOPI:2011aa}, see ~\tabref{expdataset} for additional details. Only data points with $y/y_P\leq 0.5$ have been included in the fit described in ~\secref{fitv1v2}. The same comments as for \figref{fopi1_v1rp} apply.}
\end{figure*}

\begin{figure*} [htb]
\includegraphics[width=0.925\textwidth]{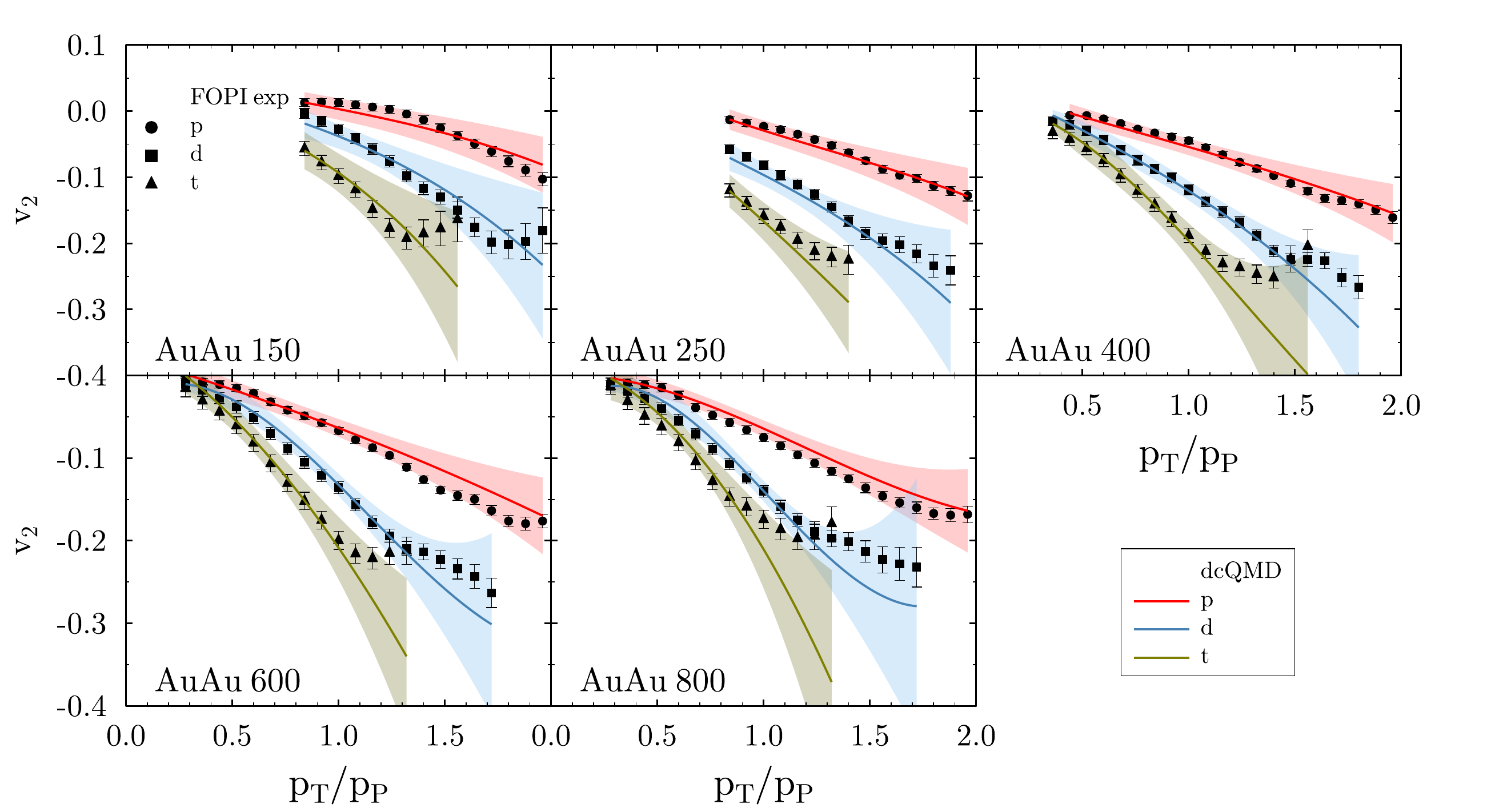}
\caption{\figlab{fopi2_v2pt} (Color online) Theoretical predictions for transverse momentum dependent $v_2$ of protons, deuterons and tritons are compared to experimental data ~\cite{FOPI:2011aa}, see ~\tabref{expdataset} for additional details. The same comments as for \figref{fopi1_v1rp} apply.}
\end{figure*}

\bibliography{references}

\end{document}